\documentclass{emulateapj}

\pdfoutput=1

\usepackage{graphicx}

\usepackage{txfonts}

\def\***#1{{\boldmath\textbf{\textsf{***#1***}}}}

\def\xip{\ensuremath{\xi_{\pi}(\pi)}}
\def\xir{\ensuremath{\xi_{p}(r_{\rm p})}}
\def\Nd{\ensuremath{N_{\rm d}}}
\def\Nr{\ensuremath{N_{\rm r}}}
\def\vmax{\ensuremath{v_{\rm max}}}
\def\Vmin{\ensuremath{V_{\rm min}}}
\def\kms{km~s$^{-1}$}
\def\ergs{erg~s$^{-1}$}
\def\Mlim{\ensuremath{M_{\rm lim}}}
\def\Msun{\ensuremath{M_{\odot}}}
\def\M200{\ensuremath{M_{180}}}
\def\fcen{\ensuremath{f}}

\def\ChBootes{\emph{Chandra}/Bo\"otes}

\begin{document}

\title{Constraining halo occupation properties of X-ray AGNs using
  clustering\\
  of \emph{Chandra} sources in the Bo\"otes survey region}


\author{S.~Starikova\altaffilmark{1,2},
R.~Cool\altaffilmark{3,4,5},
D.~Eisenstein\altaffilmark{2},
W.~Forman\altaffilmark{2},
C.~Jones\altaffilmark{2},
R.~Hickox\altaffilmark{6,7},
A.~Kenter\altaffilmark{2},
C.~Kochanek\altaffilmark{8},\\
A.~Kravtsov\altaffilmark{9},
S.~S.~Murray\altaffilmark{10,2},
A.~Vikhlinin\altaffilmark{2,11}
}
\altaffiltext{1}{Dipartimento di Astronomia, Universit\`a di
   Padova,Vicolo dell'Osservatorio 2, 35122 Padova, Italy}
 \altaffiltext{2}{Harvard-Smithsonian Center for Astrophysics, 
                  60 Garden Street, Cambridge, MA 02138}
\altaffiltext{3}{Department of Astrophysical Sciences, Princeton University, NJ 08544}
\altaffiltext{4}{Hubble Fellow}
\altaffiltext{5}{Carnegie-Princeton Fellow}
\altaffiltext{6}{Department of Physics, Durham University, South Road, Durham, DH1 3LE,
United Kingdom}
\altaffiltext{7}{STFC Postdoctoral Fellow}
\altaffiltext{8}{Dept. of Astronomy, The Ohio State University}
\altaffiltext{9}{Dept.\ of Astronomy and Astrophysics,
  Kavli Institute for Cosmological Physics, E.Fermi Institute,
  The University of Chicago, Chicago, IL 60637}
\altaffiltext{10}{John Hopkins University}
\altaffiltext{11}{Space Research Institute (IKI), Profsoyuznaya 84/32,
                 Moscow, Russia}

\shortauthors{STARIKOVA ET AL.}
\shorttitle{HALO OCCUPATION PROPERTIES OF X-RAY AGNS}

\begin{abstract}
  We present one of the most precise measurement to date of the
  spatial clustering of X-ray selected AGNs using a sample derived
  from the \emph{Chandra} X-ray Observatory survey in the Bo\"otes
  field. The real-space two-point correlation function over a redshift
  interval from $z=0.17$ to $z\sim 3$ is well described by the power
  law, $\xi(r) = (r/r_{0})^{-\gamma}$, for comoving separations
  $r\lesssim 20\,h^{-1}\,$Mpc. We find $\gamma=1.84\pm0.12$ and
  $r_{0}$ consistent with no redshift trend within the sample (varying
  between $r_{0}=5.5\pm0.6\;h^{-1}\,$Mpc for $\langle z\rangle=0.37$
  and $r_{0}=6.9\pm1.0\;h^{-1}\,$Mpc for $\langle
  z\rangle=1.28$). Further, we are able to measure the projections of
  the two-point correlation function both on the sky plane and in the
  line of sight. We use these measurements to show that the
  \ChBootes{} AGNs are predominantly located at the centers of dark
  matter halos with the circular velocity $\vmax>320\,$\kms{} or
  $\M200>4.1\times10^{12}\,h^{-1}\,M_{\odot}$, and tend to
  \emph{avoid} satellite galaxies in halos of this or higher mass. The
  halo occupation properties inferred from the clustering properties
  of \ChBootes{} AGNs --- the mass scale of the parent dark matter
  halos, the lack of significant redshift evolution of the clustering
  length, and the low satellite fraction --- are broadly consistent
  with the \cite{2006ApJS..163....1H} scenario of quasar activity
  triggered by mergers of similarly-sized galaxies.
\end{abstract}

\section{Introduction}
\label{sec:intro}

Direct observations of host galaxies of high-redshift active galactic
nuclei are hard or impossible with the current instrumentation, except
for highly obscured or low-luminosity objects. Hence, studies of the
AGN clustering properties are a unique source of information on the
AGN hosts and their environment. At low redshifts, the supermassive
black holes exist at the centers of most low-redshift galaxies, and
there is a tight correlation between the SMBH mass and the properties
of the bulges of host galaxies
\citep{1998AJ....115.2285M,2005SSRv..116..523F,2000ApJ...539L..13G}. This
suggests that most galaxies hosted an AGN at some point in their
evolution, and that AGNs at each redshift are stochastic ``markers''
of a population of galaxies in which the conditions are favorable for
accretion of matter onto the central SMBH. Through matching the
clustering properties of AGNs to those of dark matter halos, or with
those of different types of galaxies, or with those of AGNs of
different types, we can determine the typical mass scale of the AGN
hosts, their morphological type, and determine whether different types
of AGNs are hosted in the same type of objects.

Clustering of optical quasars indeed is very similar to that of
galaxies. The two-point correlation function observed at $z\sim
0.5-2.5$ in the separation range $1-20\,h^{-1}\,$Mpc is well described
by a power law, $\xi(r) = (r/r_{0})^{-\gamma}$ with a slope of
$\gamma=1.9$ and a correlation length of $r_{0}=5.5\,h^{-1}\,$Mpc
\citep{2009ApJ...697.1634R,2009ApJ...697.1656S}. Further, using the
SDSS and 2QZ quasar sample, \cite{2005MNRAS.356..415C} and
\cite{2009ApJ...697.1634R} track the evolution of the optical quasar
clustering amplitude over the redshift range $z=0.5-2.5$ and find only
a mild evolution. 

The measurements of the spatial clustering of X-ray AGNs start with
the works of \cite{2004ApJ...617..192M} and
\cite{2006ApJ...645...68Y}. The most recent measurements of the
spatial autocorrelation function of X-ray AGNs can be found in
\cite{2005A&A...430..811G,2009A&A...494...33G} and
\cite{2010ApJ...716L.209C}. In several recent works, the X-ray AGN
clustering has beed studied through cross-correlation with galaxy
catalogs
\citep{2009ApJ...696..891H,Coil:2009hg,Krumpe:2010hz}. Generally, the
two-point correlation function of the X-ray AGNs was found to be
similar to that of the optical quasars; however, the previous X-ray
studies could not constrain the evolution of the correlation function
over a sufficiently wide redshift range.

The main conclusion from previous clustering analyses is that AGNs are
located in galaxy group-sized dark matter halos \citep[$M\sim
2\times10^{12}\,h^{-1}\,M_{\odot}$, see][]{2009ApJ...697.1634R}, with
the mass scale fairly independent of the object redshift or observed
luminosity. \cite{2009ApJ...696..891H} also find some differences in
the clustering properties and color of host galaxies for the X-ray,
radio, and infrared-selected AGNs at $z=0.5$. The difference in the
AGN clustering properties and colors of their host galaxies lead
\citeauthor{2009ApJ...696..891H} to conclude that these different
techniques select distinct source populations and not simply different
stages of rapidly changing AGN properties.

In addition to determining the mass scale of the AGN host dark matter
halos, it is interesting to establish \emph{where} within the halos
the active galaxies are located. This question can be addressed by
direct observations of individual objects only in rich galaxy
clusters. Indeed, some studies
\citep{2001ApJ...548..624C,2002ApJ...573L..91M,2006ApJ...644..116M}
indicate an excess of X-ray AGNs in the cluster outskirts \citep[see,
however,][]{2010ApJ...714L.181K}. However, the majority of quasars and
AGNs are located within galaxy group-sized objects, where it is hard
or impossible to independently localize the centroid of the system,
especially at high redshifts. For such systems, it is possible to
determine the fraction of objects in satellite galaxies through
analysis of the two-point correlation function at small
separations. An example of such an approach can be found in
\cite{2009MNRAS.397.1862P}. Based on cross-correlation of the SDSS
quasar and Large Red Galaxies samples, these authors argue that a high
fraction, $>25\%$, of the optical quasars must be located in
non-central galaxies.

One of the main goals of the present work is to constrain the location
of X-ray emitting AGNs within their host dark matter halos by a more
direct method. The effect we are using is the strong dependence of a
galaxy's peculiar velocity on its location within the host halo. The
central galaxies are predicted \citep[and observed, at least in
massive clusters, see][]{2001AJ....122.2858O} to be nearly at rest
with respect to the host halo, and their random motions correspond to
motions of the halos as a whole. The satellite galaxies move at
approximately the virial velocities \emph{within} their parent halo. As a
result, the satellite galaxies have much faster peculiar motions and
form ``finger of God'' structures in the radial velocity space. The
effect of high peculiar velocities of the satellite galaxies can be
detected through comparison of the objects' clustering properties as a
function of projected separation, $r_{p}$, and the line-of-sight
separation, $\pi$. Peculiar motions do not affect $r_{p}$ but can
strongly distort $\pi$ because it is derived from the measured
redshift.

Using peculiar motions in the previous studies of the AGN clustering
was hard because of the small number of objects and insufficient
accuracy of the redshift measurements\footnote{For example, the
  redshift uncertainties corresponded to peculiar velocities of
  $\approx 420-500\,$\kms{} in the 2QZ sample
  \citep{2005MNRAS.356..415C}. With such large uncertainties,
  \citeauthor{2005MNRAS.356..415C} could not extract useful
  information from the velocity-space distrortions although they were
  included in the modeling of the spatial correlation
  function.}. Fortunately, we now can use an excellent sample for such
studies. The \emph{Chandra} survey of 9.3~deg$^2$ in the Bo\"otes
region \citep{2005ApJS..161....1M} provides a uniformly selected
sample of $>3000$ X-ray selected AGNs, for a uniform subsample of
which ($\sim 1900$ sources) high-quality spectra were measured with
MMT/Hectospec (Kochanek et al., in preparation).

\begin{figure}
  \centerline{\includegraphics[width=0.99\linewidth]{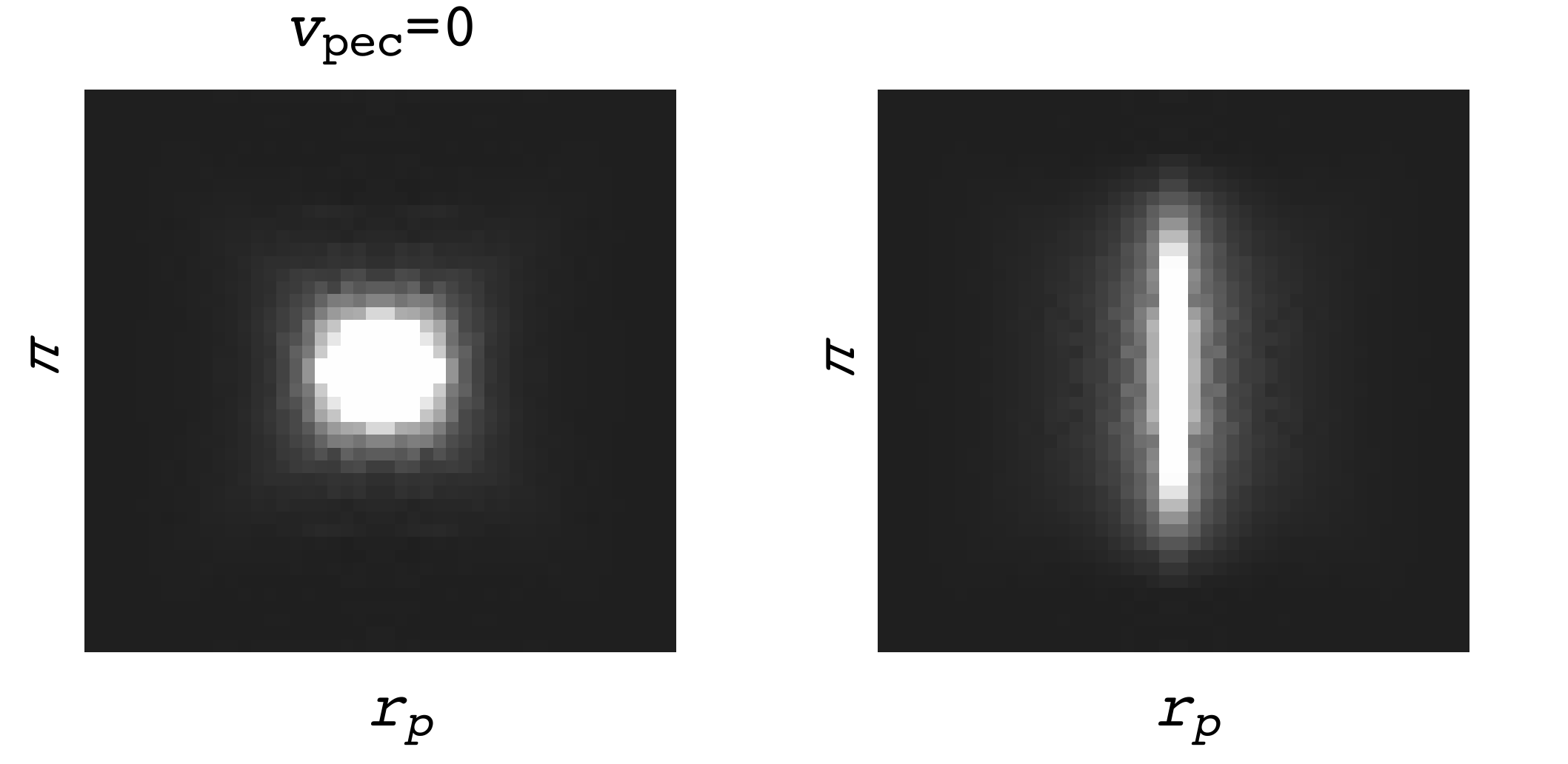}}
  \caption{The two-dimensional correlation function $\xi(r_{\rm p},
    \pi)$ for objects located within dark matter halos with the
    maximum circular velocity $V_{\rm max}>250 \,\,{\rm km \,\,
      s^{-1}}$ at $z=1$, plotted as a function of transverse ($r_{\rm
      p}$) and line-of-sight ($\pi$) pair separation. 70\% of objects
    are forced to be at the centers of such halos, while 30\% are put
    in one of the satellite subhalos. In the left panel, the peculiar
    velocities were set to 0.}
\label{fig:2Dksi}
\end{figure}

Full information on the object clustering and peculiar velocities is
contained in the two-dimensional correlation function, $\xi(r_{\rm
  p},\pi)$. An example derived from the numerical simulations we use
in this Paper is shown in Fig.~\ref{fig:2Dksi}. See, e.g.,
\cite{2002ApJ...571..172Z} for an example of the detailed modeling of
the $\xi(r_{\rm p},\pi)$ function measured for the SDSS
galaxies. Unfortunately, such a detailed modeling is impossible for
the present high-$z$ AGN samples due to limited statistics. However,
we show that useful information can still be obtained using
projections of $\xi(r_{\rm p},\pi)$ on the radial velocity direction
and on the sky plane, $\xip$ and $\xir$, respectively (formally
defined in \S~\ref{ssec:fun:definitions} below). Using numerical
cosmological simulations, we show that if a substantial fraction of
objects lie in satellite galaxies, $\xip$ is expected to be
significantly in excess of $\xir$ in the range of comoving distances
$1-10\,h^{-1}$\,Mpc (\S~\ref{sec:model:corr:functions}). The observed
correlation functions do not show such an excess
(\S~\ref{sec:res:vmin-f}), which we exploit to put an upper limit on
the fraction of X-ray AGNs in the satellite galaxies.
Finally, we explore the redshift evolution of the clustering length
and compute the typical mass scale of the AGN host dark matter halos
and put constraints on the AGN duty cycle.

All cosmology-dependent quantities are computed assuming a spatially
flat model with parameters $\Omega_{M}=0.268$ and
$\Omega_{\Lambda}=0.732$ \citep[best-fit $\Lambda$CDM paramaters
obtained from the combination of CMB, supernovae, BAO, and galaxy
cluster data, see][]{2009ApJ...692.1060V}. All distances are comoving
and given with explicit $h$-scaling, where the Hubble constant is
$H_0=100\,h^{-1}\,$km~s$^{-1}$~Mpc$^{-1}$. The parameter uncertainties
are quoted at a confidence level of 68\%.

\begin{figure}
\centering
\includegraphics[width=1.0\linewidth]{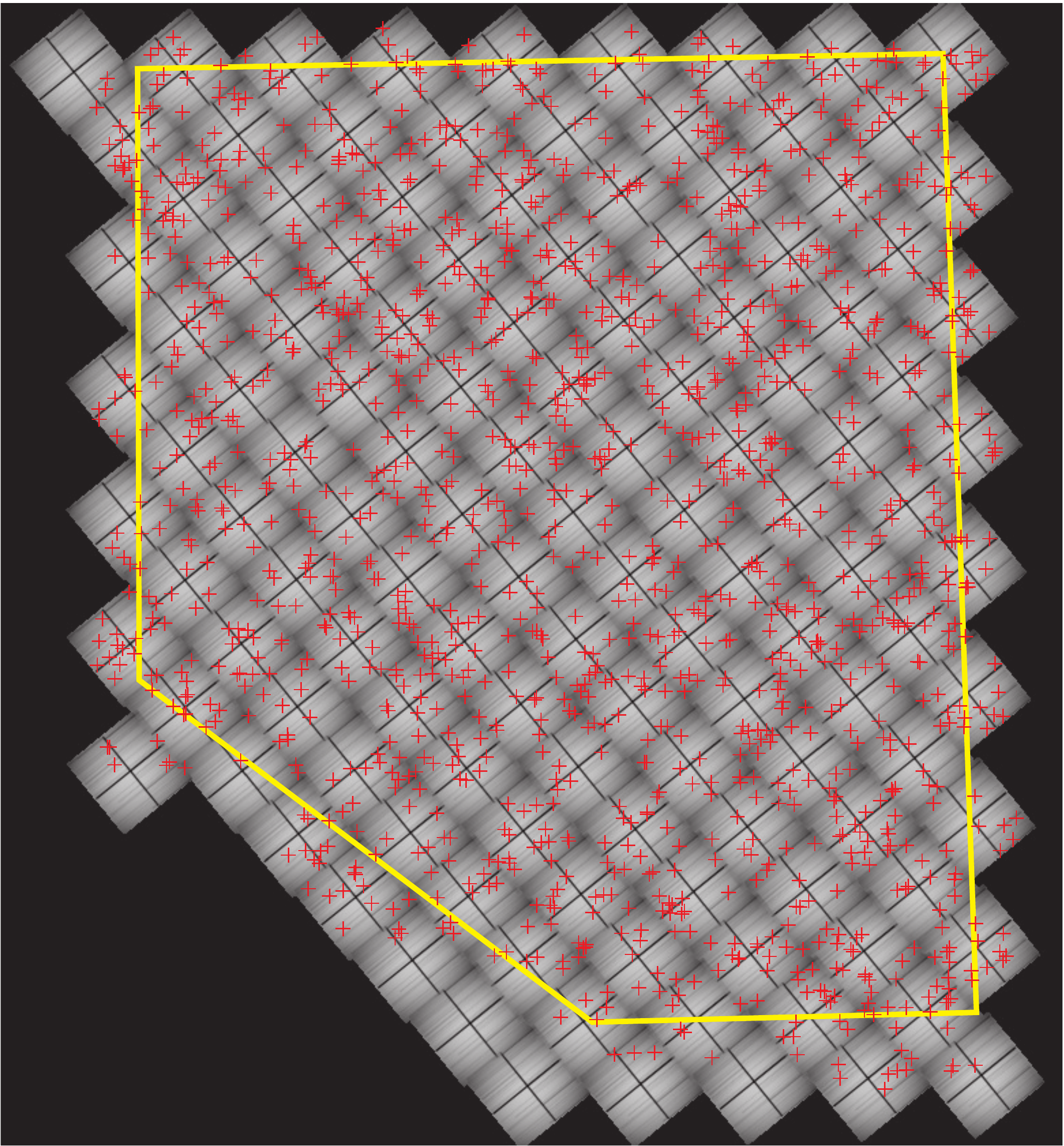}
\caption{\emph{Chandra} survey in the Bo\"otes field. Grey scale map
  shows the combined \emph{Chandra} sensitivity map (exposure maps
  for 126 individual pointings, each multiplied by the point
  source detection sensitivity as a function of off-axis distance
  derived in \S\,\ref{sec:exposure-map}). Crosses mark the location of
  X-ray sources used in this work. The polygon shows conservative
  boundaries of the spectroscopic survey. We consider only 1282
  sources within the polygon boundaries in the clustering analysis.}
\label{fig:field}
\end{figure}

\section{AGN Sample}
\label{sec:AGN}

We use a sample of high-$z$ AGNs derived from the \emph{Chandra} X-ray
survey in the 9.3~deg$^{2}$ Bootes field of the NOAO Deep Wide-Field
Survey \citep{2005ApJS..161....1M}. The region was uniformly covered
with a grid of overlapping 5~ksec ACIS-I pointings providing a
sensitivity of $4.7\times10^{-15}$\,erg~s$^{-1}$~cm$^{-2}$ in the
0.5--2~keV energy band \citep{2005ApJS..161....9K}. Extensive optical
data exist for this field and 98\% of X-ray sources have optical or
infrared counterparts \citep{2006ApJ...641..140B}. The redshifts for
X-ray sources with optical counterparts brighter than $I=21.5$ were
uniformly obtained with the MMT/Hectospec instrument in the AGN and
Galaxy Evolution Survey (AGES, C. S. Kochanek et al., in
preparation). Further details on the X-ray and optical observations
can be found in \cite{2009ApJ...696..891H}.

\begin{figure}
\centering
\includegraphics[width=1.0\linewidth]{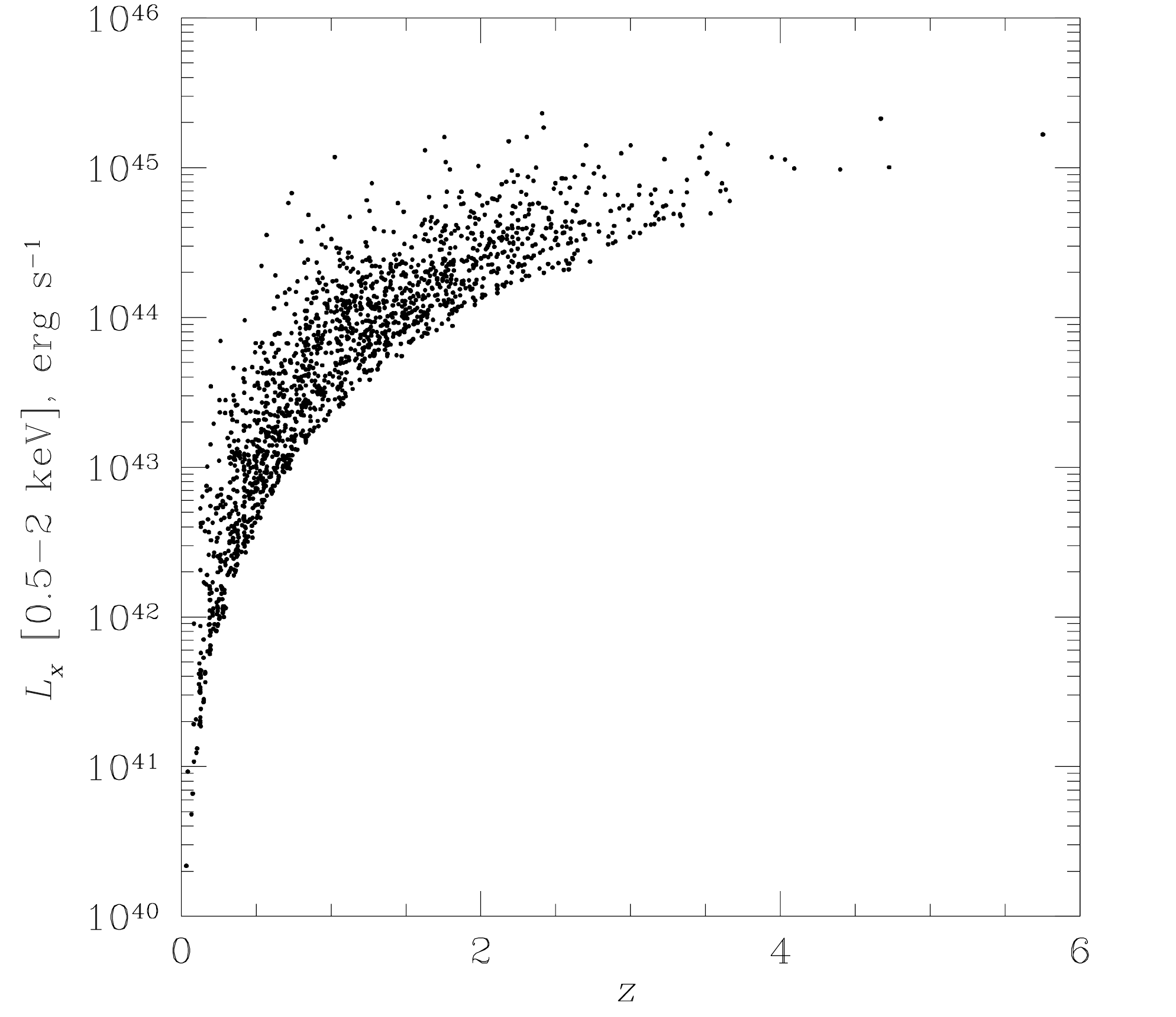}
\caption{The redshift-luminosity diagram for \emph{Chandra} sources in
  the Bo\"otes field used in our clustering analysis.}
\label{fig:lx-z}
\end{figure}

For clustering analysis, we use 1282 X-ray selected AGNs with
spectroscopic redshifts. Redshift measurements effectively introduce
an additional selection criterion for the sample, based on the optical
magnitude of the counterparts, $I<21.5$. This selection removes $\sim
35\%$ of the X-ray sources and likely introduces a high-$z$ cutoff in
the redshift distribution of our sources
(\S~\ref{sec:dNdz}). Fortunately, the redshift measurements were done
with Hectospec \citep{2005PASP..117.1411F} in several passes, which
excludes the so called ``fiber collisions'' problem. The locations of
X-ray sources with spectroscopic redshifts are marked in
Fig.~\ref{fig:field}, and their luminosity-redshift diagram is shown
in Fig.~\ref{fig:lx-z}.

To reconstruct the correlation function of sources, we need to
simulate catalogs of randomly distributed sources whose distribution
follows the sensitivity variations and the redshift distribution of
our catalog. Below, we describe how these functions were derived from
our \emph{Chandra} sample.

\subsection{Spatial variations of sensitivity}
\label{sec:exposure-map}

\emph{Chandra} sensitivity for the detection of point X-ray sources is
not uniform across the field of view. Sensitivity variations ($\sim
\pm 25\%$, see below) are imprinted on the distribution of detected
sources. The typical spatial scale of the sensitivity variations is
several arcmin, which corresponds to the comoving distance of
$1-2\,h^{-1}$~Mpc at $z=1$. This is comparable to the distance scale
where we measure clustering, therefore these variations must be taken
into account.

The sensitivity variations are caused mainly by two effects --- 1) the
vignetting of the \emph{Chandra} X-ray telescopes, and 2) the
degradation of the Point Spread Function (PSF) away from the optical
axis. The mirror vignetting is well-calibrated and its effect can be
computed for a given source population. However, the effects of PSF
degradation on the source detection efficiency in the low-photon
regime are very complex. Therefore, it is best to measure the combined
effect of the sensitivity variations empirically.

\begin{figure}
\centering
\centerline{\includegraphics[width=0.99\linewidth]{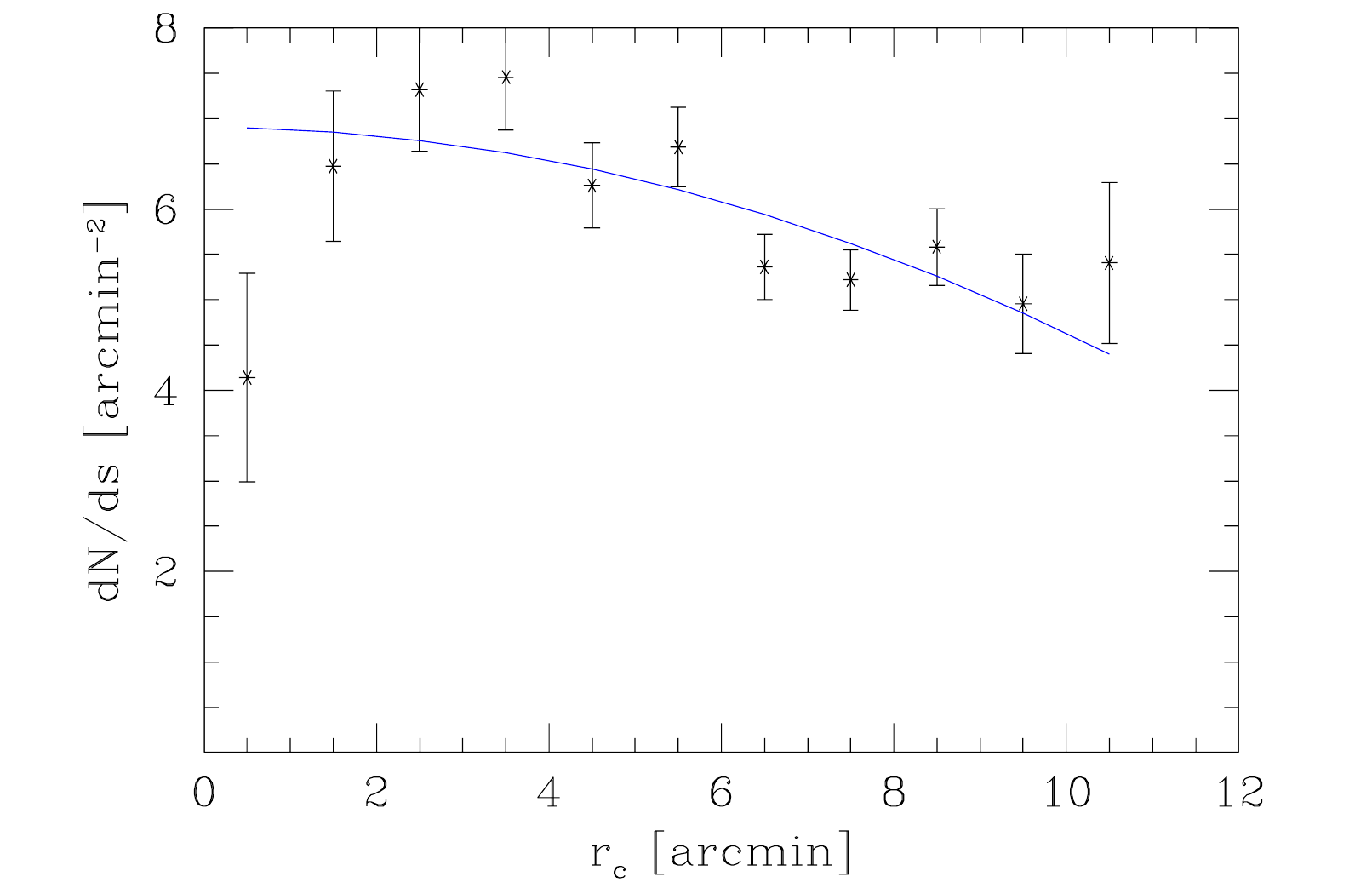}}
\caption{The observed surface density of detected X-ray sources as a
  function of the distance from the telescope optical axis. The sharp
  decrease at $r_{c}<1'$ is attributed to the gap between the
  \emph{Chandra} ACIS-I CCDs. The gradual decline at larger radii
  reflects the sensitivity variation caused by mirror vignetting and
  PSF degradation. The solid line is the best-fit second-order
  polynomial (eq.~[\ref{frm:sf}]).}
\label{fig:sf}
\end{figure}

Both the mirror vignetting and PSF degradation are approximately
azimuthally symmetric. Therefore, we can assume that the detection
sensitivity is a function of the source offaxis angle, $r_{c}$ and
measure it from the radial profile of the surface density distribution
of detected X-ray sources, averaged over all the 126 fields. The
results are shown in Fig.~\ref{fig:sf}. For a uniform sensitivity, we
would expect a constant surface density, while in reality we find that
the average surface density is $\sim 7$~arcmin$^{-2}$ near the optical
axis, and falls to $\sim 5$~arcmin$^{-2}$ at off-axis distances of
10~arcmin.  The derived radial profile of the source surface density
can be modeled with a second-order polynomial,
\begin{equation}
  \frac{dN}{ds}=6.9\left(1-(r_{c}/17.4^{\prime})^2\right).
  \label{frm:sf}
\end{equation}

In addition to this gradual variation with radius, there are sharp
features in the spatial sensitivity pattern related to the gaps
between the \emph{ACIS-I} CCDs. In particular, these gaps are
responsible for the drop in the number of detected sources at $r_{\rm
  c}<1'$ in Fig.~\ref{fig:sf}. These sensitivity variations can be
adequately taken into account using the standard \emph{Chandra}
exposure maps. In doing so, we assume that the expected surface density
of X-ray sources at a given location is proportional to the effective
exposure at this location. This assumption is justified because the
source detection in the \ChBootes\ survey is well in the
photon-limited regime (the background is unimportant) and the observed
cumulative source counts are very close to $N(>S) \propto S^{-1}$
around our flux limits \citep{1993A&A...275....1H,1995ApJ...451..553V}.

Our final sensitivity map (Fig.~\ref{fig:field}) consists of the
merged set of \emph{Chandra} exposure maps computed for the individual
126 pointings, each multiplied by the radial sensitivity pattern given
by eq.~(\ref{frm:sf}). This sensitivity map is taken into account in
the derivation of the correlation function through the generation of
the appropriate catalogs of random sources. The total area within the
conservative boundaries of the spectroscopically surveyed region of
the Bootes field is 7.30~deg$^{2}$. The total effective area, taking
into account the gaps between the \emph{Chandra} CCDs and degradation
of the detection efficiency, is 5.90~deg$^{2}$.

\begin{figure*}
\vspace*{-7mm}
\centerline{%
\includegraphics[width=0.99\columnwidth]{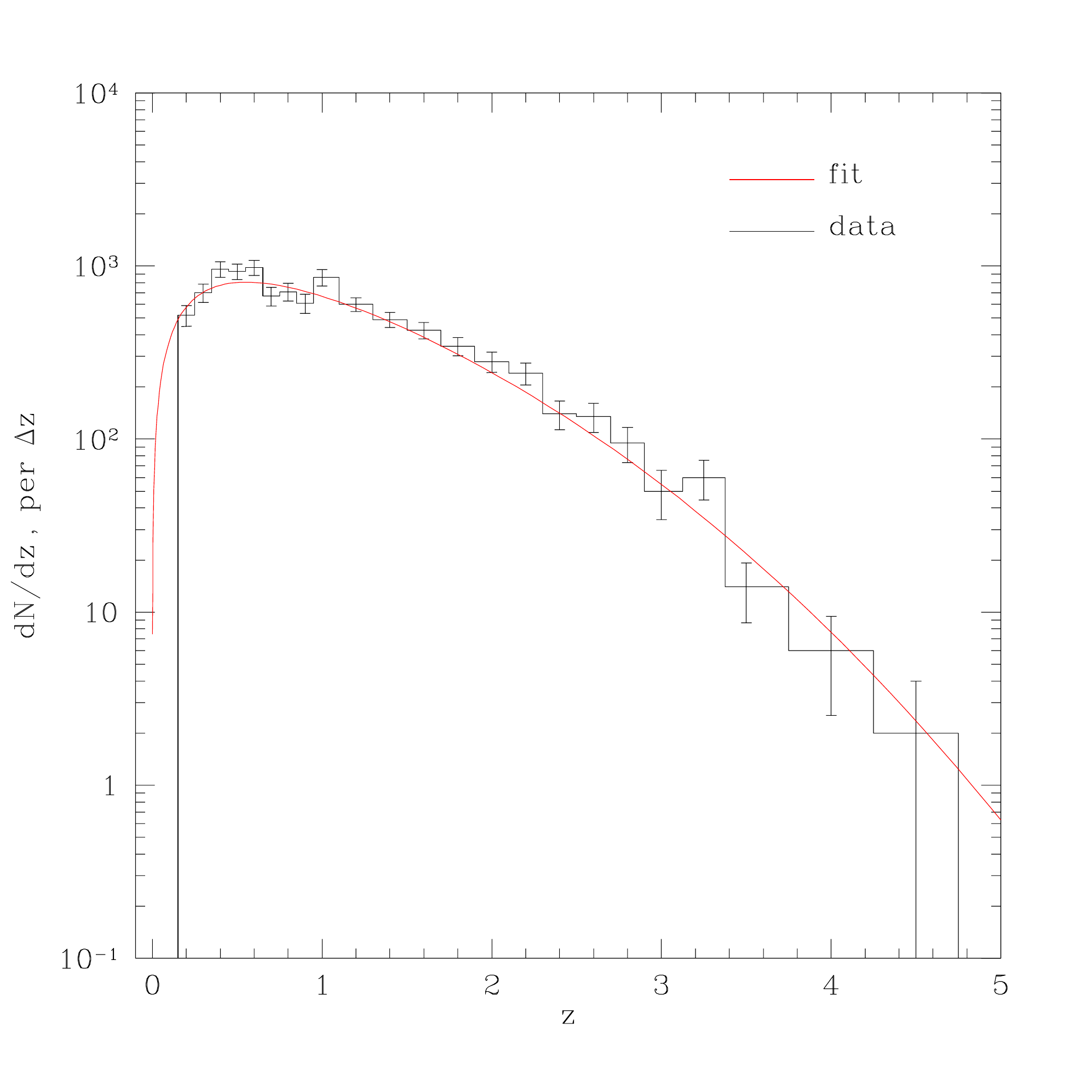}
\includegraphics[width=0.99\columnwidth]{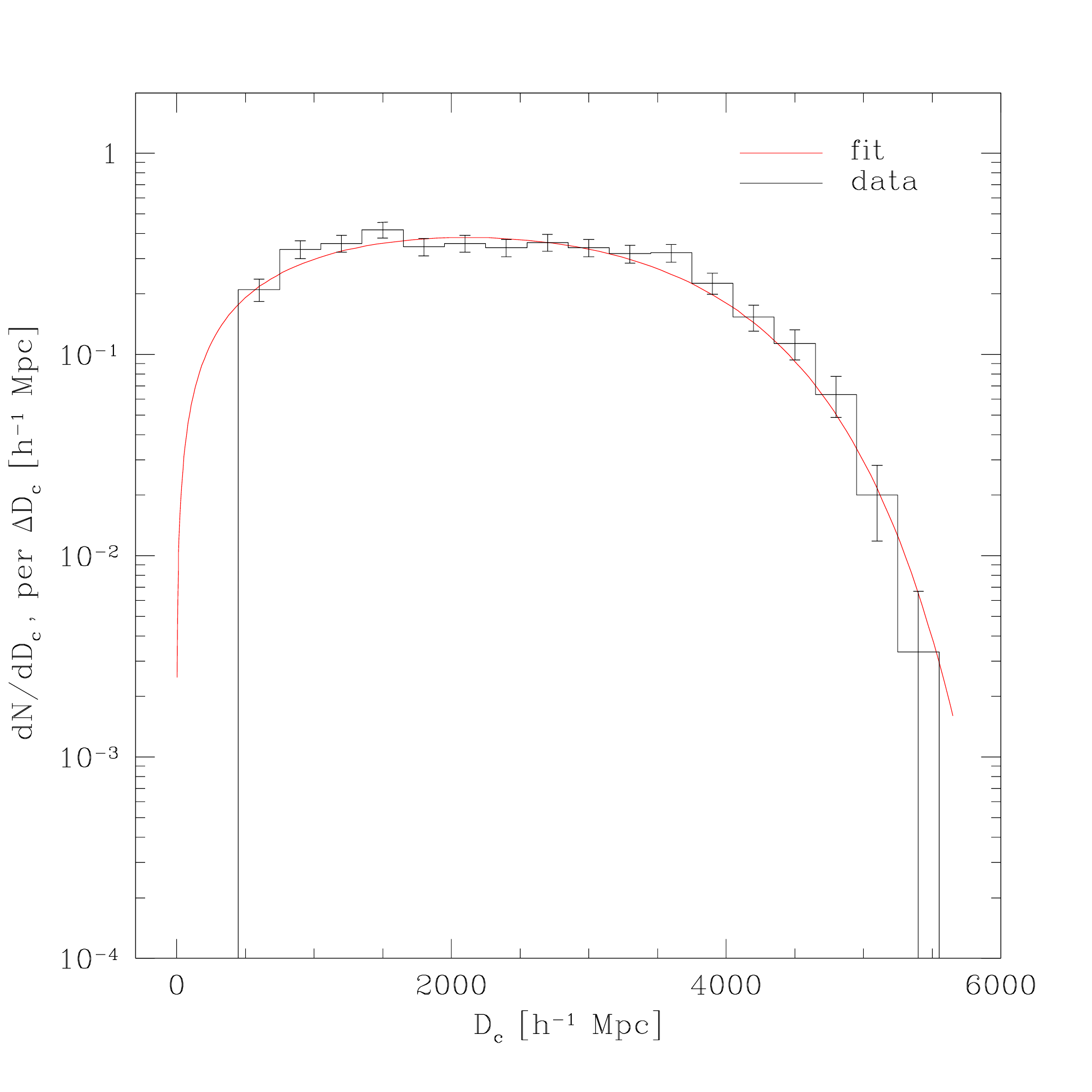}
}
\vspace*{-3mm}
\caption{The observed distribution of \ChBootes{} AGNs as a function
  of redshift \emph{(left)} and comoving distance \emph{(right)}. The
  best fit by eq.~[\ref{frm:dndz}] is shown by solid lines.}
\label{fig:dndz}
\end{figure*}

\subsection{Redshift distribution}
\label{sec:dNdz}

The model for the source redshift distribution, $dN/dz$, should
reflect both the intrinsic variations of the comoving number density
with redshift and all selection effects of the catalog. A commonly
used approach is to model the observed $dN/dz$ distribution with a
high-order polynomial \citep[see,
e.g.,][for recent
examples]{2005MNRAS.356..415C,2009ApJ...697.1634R}. This approach
works well for catalogs with a large number of sources. However, for
smaller catalogs, like ours, there is a danger that a high-order
polynomial fit will follow statistical fluctuations in the observed
$dN/dz$, while a low-order polynomial would be unable to adequately
model the strong gradients at low $z$. Therefore, we fit the
redshift distribution of AGNs in the Bo\"otes field with a parametric
model based on several physical assumptions.

The first component of the model represents the cosmological comoving
volume per unit redshift, $dN_{1}/dz\propto dV/dz$. The second
component is a power law function of the minimum luminosity which
corresponds to the \emph{Chandra} flux limit at redshift $z$,
$dN_{2}/dz\propto L_{\rm min}^{\alpha/2}\propto d_{L}^{\alpha}$, where
$d_{L}$ is the luminosity distance. This component represents the
effect of the low-$L_{x}$ cutoff of the intrinsic luminosity function
introduced by the selection which is primarily based on \emph{Chandra}
detections. It also can describe the evolution of the luminosity
function at high $z$. The third component is a high-$z$ cutoff modeled
by a broad Gaussian, $dN_{3}/dz\propto
\exp\left(-d_{L}^{2}/C^{2}\right)$. This component can represent the
high-$L_{x}$ cutoff or steepening of the intrinsic AGN luminosity
function, and also can describe various observational limits
implicitly built into our catalog (e.g., a lower efficiency of optical
identifications and redshift measurements for the highest-$z$ X-ray
sources). This simple analytic model,
\begin{equation}
  \label{frm:dndz}
  \frac{dN}{dz}=\mathrm{const}\times\frac{dV}{dz}\,d_{L}^{\alpha}\,\exp\left(-\left(d_{L}/C\right)^2\right),
\end{equation}
which has only two free parameters provides a strikingly good fit to
the observed redshift distribution of the Bo\"otes X-ray selected AGNs
(Fig.~\ref{fig:dndz}). The best-fit values are $\alpha=-1.07$ and
$C=1.50\times10^{3}\,h^{-1}\,$Mpc. We use the arguments outlined above
only as a motivation for a good analytical description of the $dN/dz$
distribution for our sources. The functional form and derived
parameters are not meant to represent the true X-ray luminosity
function or its evolution.

Figure~\ref{fig:dndz} also demonstrates the general characteristics of
our sample. The peak in the observed $dN/dz$ distribution is near
$z\approx 0.6$. The median redshift of the sample is $z_{\rm
  med}=1.04$. The tail in the redshift distribution extends to
$z\approx4.5$ but the fraction of AGNs with $z>3$ is very
small. Overall, the clustering properties of sources in our sample are
most sensitive to the distribution of the X-ray AGN population near
$z\approx 1$.
 
\section{Two-point Correlation Function of \ChBootes{} AGNs}
\label{sec:function}
\subsection{Definitions}
\label{ssec:fun:definitions}

Because the volume is never sampled completely in astronomical
surveys, the derivation of the two-point correlation function from the
data uses mock catalogs of intrinsically randomly distributed objects,
which faithfully reproduces all observational distortions introduced
by the survey. Examples of such distortions are boundaries of the
survey region, gaps in the data or spatial variations of the
sensitivity, variations of the selection efficiency with redshift,
etc. Given the catalog of observed sources and the mock random
catalog, the two-point correlation function can be estimated
\citep{LS93} as
\begin{equation}
\label{frm:LSE}
\xi=\frac{\Nr(\Nr-1)}{\Nd(\Nd-1)}\frac{DD}{RR}-\frac{\Nr-1}{\Nd}\frac{DR}{RR}+1,
\end{equation}
where $DD$ is the number of source pairs in the data for the given
distance interval, $RR$ is the corresponding number of pairs in the
random catalog, $DR$ is the number of pairs between the data and random
catalog, and $N_{d}$ and $N_{r}$ are the numbers of objects in the data
and random catalogs, respectively. Statistical uncertainties for $\xi$
can be estimated as
\begin{equation}
\delta\xi=\frac{1+\xi}{\sqrt{DD}}
\label{frm:poisson error}
\end{equation}
\citep{Peebles73}; this equation includes both the Poissonian shot
noise and intrinsic variance terms. To verify the accuracy of the
error by eq.~\ref{frm:poisson error}, we used the sample varience of
the correlation functions measured in the mock catalogs derived from
the Millenium simulation \citep{2007MNRAS.376....2K} for the survey
geometry and object properties similar to those of the \ChBootes{}
survey. This analysis showed that eq.~\ref{frm:poisson error} is
accurate at tsmall scales but may underestimate the uncertainties at
large scales. The correction factor can be described by a smooth
function which is 3\%, 23\%, and 42\% at separations of $1$, $6$, and
$15\,h^{-1}\,$Mpc, respectively. This correction is applied to the
statistical uncertainties estimated by eq.~\ref{frm:poisson error}.

The correlation function in real space is expected to be isotropic, so
$\xi$ is a function of the 3D separation only. When the object
redshifts are used to derive the distances, the correlation function
is distorted in the line-of-sight direction because of large-scale
flows (the Kaiser \citeyear{1987MNRAS.227....1K} effect) and ``fingers
of God'' arising within the virialized dark matter halos. The
correlation function should then be measured as a function of the
projected separation, $r_{p}$, and the line-of-sight separation,
$\pi$. Equations~[\ref{frm:LSE}] and~[\ref{frm:poisson error}] still
can be used, but the pairs must be counted for each combination
$(r_{p},\pi)$.

Given the angular separation between two objects, $\theta$, and
redshifts, $z_{1}$ and $z_{2}$, the comoving separations $r_{p}$ and
$\pi$ can be computed as follows. First, one computes the radial
comoving distances, $D_{c,1}$ and $D_{c,2}$, corresponding to the
object redshifts \citep[see, e.g.,][]{1999astro.ph..5116H}. Then,
following \cite{DP83} we have
\begin{equation}
  \label{eq:pi}
  \pi=|D_{c,1}-D_{c,2}|,
\end{equation}
\begin{equation}
  \label{eq:r_p}
  r_{p}=\left[2D_{c,1}D_{c,2}(1-\cos\theta)\right]^{1/2}.
\end{equation}
One can also define a formal 3D separation,
\begin{equation}
  s=\left(r_{p}^{2}+\pi^{2}\right)^{1/2},
\end{equation}
but it should be kept in mind that $s$ is not equivalent to the true
3D separation, $r$, because of the redshift space distortions.

As only the line-of-sight separations, $\pi$, are affected by the object peculiar
velocities, it is useful to consider the correlation function
projected on the sky plane,
\begin{equation}\label{eq:w(r_p):1}
  w_{p}(r_{p})= \int_{-\infty}^{\infty}{\xi\left(\sqrt{r_{p}^2+\pi^2}\right)\,d\pi},
\end{equation}
because it is not modified by the redshift-space distortions \citep{DP83}:
\begin{equation}\label{eq:w(r_p):2}
  \int_{-\infty}^{\infty}\xi^{\rm (true)}\left(\sqrt{r_{p}^2+\pi^2}\right)\,d\pi
  = \int_{-\infty}^{\infty}{\xi^{\rm (obs)}(r_{p},\pi)\,d\pi}.
\end{equation}

\begin{figure*}
\vspace*{-7mm}
\centerline{\includegraphics[width=0.99\columnwidth]{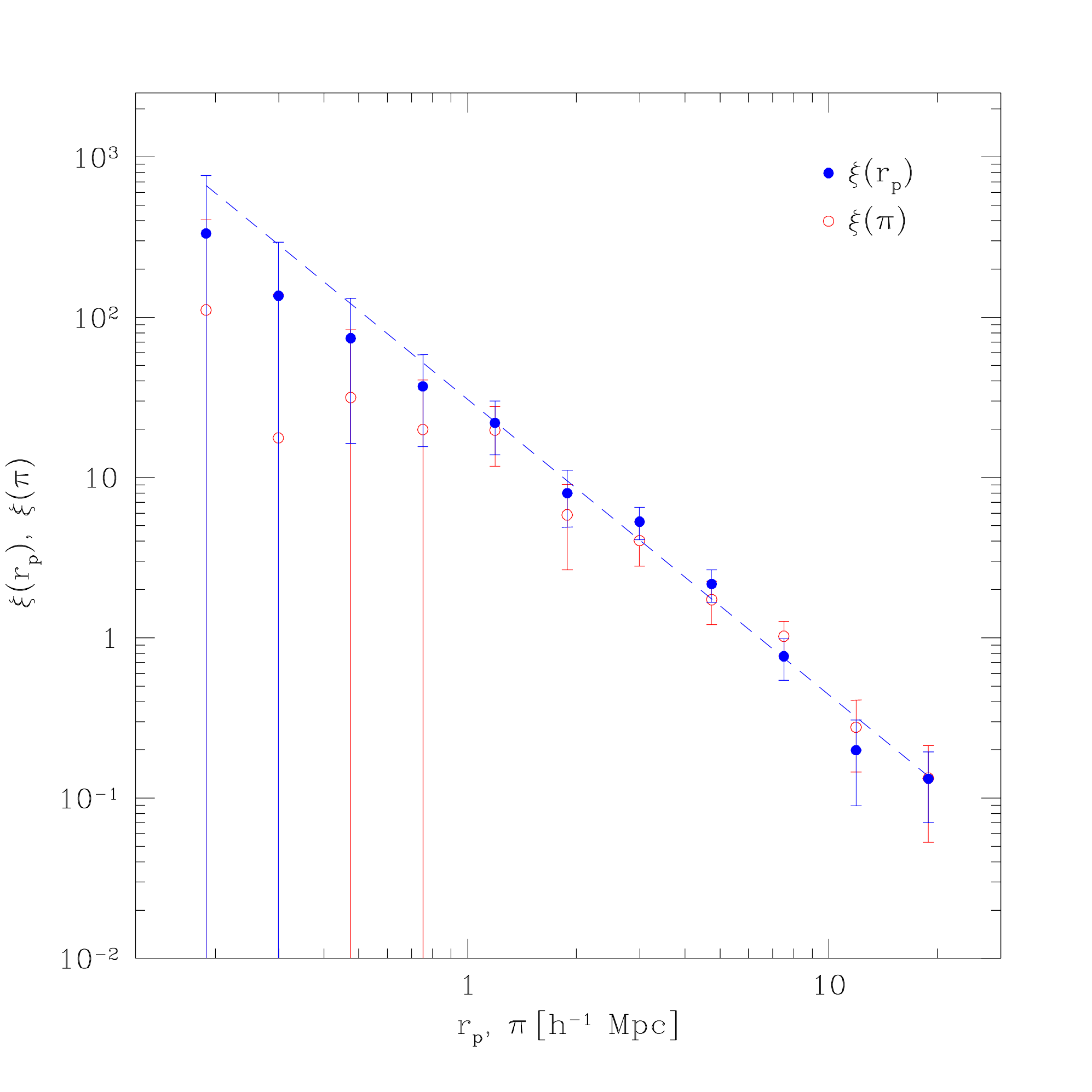}
\includegraphics[width=0.99\columnwidth]{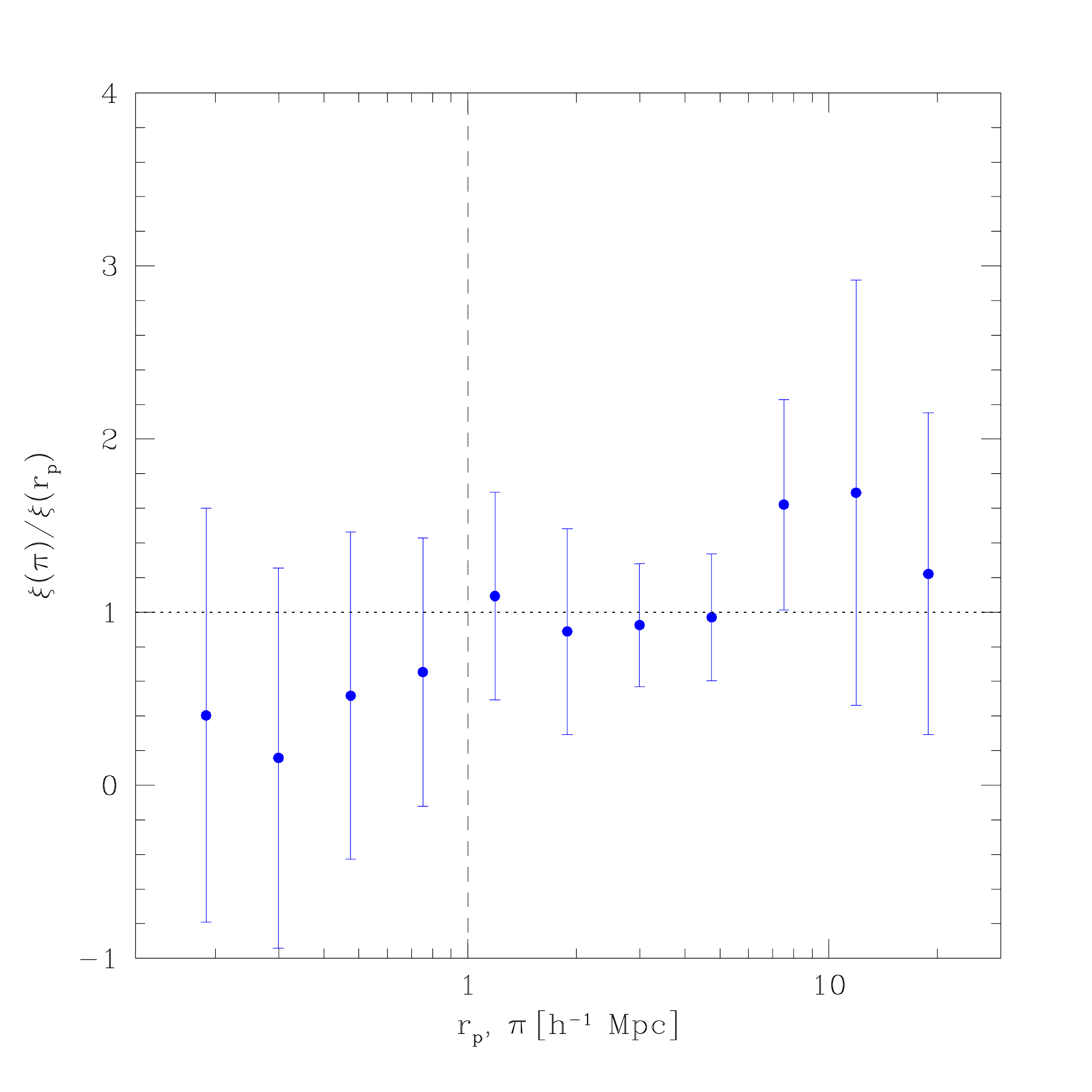}}
\vspace*{-3mm}
\caption{\emph{(a)} The two-point correlation functions of
  Bootes/\emph{Chandra} AGNs estimated from projections on the sky
  plane ($\xir$, filled circles) and on the line of sight ($\xip$,
  open circles). See \S\,\ref{ssec:fun:results} for the description of
  our procedure to derive $\xir$ and $\xip$. Dashed line shows a
  power law fit to the $\xir$ data. \emph{(b)} The ratio of two
  projected functions. At small separation
  ($\pi\lesssim1\,h^{-1}\,$Mpc, indicated by the vertical dashed
  line), the ratio can be affected by uncertainties of the AGES
  redshift measurements. At larger separations, we observe no
  statistically significant difference between $\xir$ and $\xip$.}
\label{fig:ksi}
\end{figure*}

Because of the insensitivity of $w_{p}(r_{p})$ to the redshift-space
distortions, most of the studies which involve detailed modeling of
the shape of the galaxy two-point correlation function are based on
fitting the $w_{p}(r_{p})$ measurements
\citep[e.g.,][]{2005ApJ...621...22Z,2009MNRAS.397.1862P}. Using
$w_{p}(r_{p})$ instead of $\xi(r)$ is particularly straightforward in
those cases when $\xi(r)$ can be sufficiently accurately approximated
by a power law, $\xi(r)=(r/r_{0})^{-\gamma}$. It follows from
eq.~[\ref{eq:w(r_p):1}] that in this case the relation between $\xi$
and $w_{p}(r_{p})$ is simply \citep{peebles80}
\begin{equation}
  \label{eq:w(r_p):for:power:law}
  w_{p}(r_{p})=A(\gamma)\,r_{p}\,(r_{p}/r_{0})^{-\gamma}, 
\end{equation}
where
\begin{equation}\label{eq:A(gamma)}
  A(\gamma)=\Gamma(1/2)\Gamma[(\gamma-1)/2]/\Gamma(\gamma/2).
\end{equation}
Therefore, the correlation length, $r_{0}$, and the slope of the true
3D correlation function $\xi$ can be obtained immediately from the
power-law fit to $w_{p}(r_{p})$.

Below, we also use a projection of the measured 3D correlation
function on the line-of-sight direction,
\begin{equation}\label{eq:w(pi)}
  w_{\pi}(\pi)=\int_{-\infty}^{\infty}{\xi^{\rm (obs)}(r_{p},\pi)\,dr_{p}}.
\end{equation}
In the absence of peculiar motions, $w_{\pi}(\pi)$ should be equivalent to
$w_{p}(r_{p})$. In particular, for a power-law $\xi(r)$,
eq.~[\ref{eq:w(r_p):for:power:law}] will be valid also for $w_{\pi}(\pi)$.

Large separations do not contribute significantly to the integrals in
equations~\ref{eq:w(r_p):1} and~\ref{eq:w(r_p):2} but add noise,
therefore in practice the integration is truncated at some finite
separation. A good choice for studies of AGN samples is to truncate
the integration at $\approx\pm40\,h^{-1}\,$Mpc \citep[see,
e.g.,][]{2009A&A...494...33G}. Following this and other works we
investigated how the derived correlation length depends on the choice
of the cutoff distance (see, e.g., Fig.~9 in
\citeauthor{2005A&A...430..811G}~\citeyear{2005A&A...430..811G} or
Fig.3 in \citeauthor{Allevato:2011uz}~\citeyear{Allevato:2011uz}). For
our sample, the convergence is reached at a cutoff radius of
$20\,h^{-1}\,$Mpc (for cutoffs between 20 and $60\,h^{-1}\,$Mpc, the
derived $r_{0}$ varies by less than 4\%). Based on this analysis, we
trancated the integration at comoving separations $\pm
30\,h^{-1}\,$Mpc, and further reduced the sensitivity of our results
to this choice by applying small corrections to the integrated
functions. Assuming that the true correlation function follows a power
law with index $\gamma$ at large separations, we can calculate the
effect of this truncation as
\begin{equation}
  \label{eq:cgamma(rp)}
  C_{\gamma}(r)=\frac{\int_{0}^{30}\left(r^2+l^2\right)^{-\gamma/2}\,dl}
  {\int_{0}^{\infty}\left(r^2+l^2\right)^{-\gamma/2}\,dl}.
\end{equation}
The correction coefficient, $C_{\gamma}(r)$, is close to 1 for
$r\ll30\,h^{-1}\,$Mpc and gradually decreases at larger distances. For
$\gamma=2$ (close to our best-fit value), $C_{\gamma}=0.96$ at
$r=2\,h^{-1}\,$Mpc, 0.87 at $r=6\,h^{-1}\,$Mpc ($\approx$ the
correlation length), and 0.63 at $r=20\,h^{-1}\,$Mpc ($\approx$ the
largest separation at which the projected correlation is still
marginally detectable).

Using equations~\ref{eq:w(r_p):for:power:law} and~\ref{eq:cgamma(rp)},
we can define the quantities
\begin{equation}
  \label{eq:xi(r_p)}
  \xir \equiv \frac{1}{A(\gamma)\,C_{\gamma}(r_{p})}\frac{w_{p}(r_{p})}{r_{p}},
\end{equation}
\begin{equation}
  \label{eq:x(pi)}
  \xip \equiv \frac{1}{A(\gamma)\,C_{\gamma}(\pi)}\frac{w_{\pi}(\pi)}{\pi},
\end{equation}
where $\gamma$ is determined from the power-law fit to the measured
$w_{p}(r_{p})$ (eq.~[\ref{eq:w(r_p):for:power:law}]). Thus defined, $\xir$
should be a close approximation to the true 3D correlation function
$\xi(r)$. In the absence of redshift-space distortions, $\xip$ also
should approximate $\xi(r)$. However, the peculiar motions (and in
particular, the ``finger of God'' effect) suppress $\xip$ on the
smallest separations and enhance it relative to $\xi(r)$ on
intermediate scales (see below). Therefore, $\xir$ can be used to
determine the correlation length of our objects and hence the mass
scale of their host dark matter halos. The ratio $\xip/\xir$ reflects
the amplitude of the peculiar motions and hence can be used to
constrain the fraction of objects located in the satellite dark matter
subhalos.

\subsection{Results}
\label{ssec:fun:results}

The correlation functions $\xir$ and $\xip$ derived for our complete
sample are shown in Fig.~\ref{fig:ksi}a (filled and open circles,
respectively). Technically, we derive the functions as follows. First,
we estimate the two-dimensional correlation function,
$\xi(r_{p},\pi)$, on a grid of separations with equal logarithmic
width for each cell, $\Delta\log d=0.2$. We use the Landy~\&~Szalay
estimator (eq.~\ref{frm:LSE}). The random catalog is generated using
the spatial sensitivity map and the model for the redshift
distribution in our sample (\S~\ref{sec:exposure-map}
and~\ref{sec:dNdz}). To minimize the additional noise, the number of
objects in the random catalog is a factor of 100~larger than that for
the AGN sample. The derived $\xi(r_{p},\pi)$ is then integrated in the
radial and sky plane directions to obtain $w_{p}(r_{p})$ and
$w_{\pi}(\pi)$ (eq.~\ref{eq:w(r_p):2} and~\ref{eq:w(pi)}). As
discussed in \S~\ref{ssec:fun:definitions}, the integration is limited
to separations $\pm30\,h^{-1}\,$Mpc to minimize noise. We then fit a
power law (eq.~\ref{eq:w(r_p):for:power:law}) to $w_{p}(r_{p})$ to
measure the slope of the correlation function,
$\gamma=1.84\pm0.12$. The correlation length measured for the full
sample is $r_0=6.41\pm0.44\,h^{-1}\,$Mpc.  We then compute the
renormalization factor $A(\gamma)$ (eq.~\ref{eq:A(gamma)}) and correct
for the truncation of the integration in $w_{p}(r_{p})$ and
$w_{\pi}(\pi)$ at $30\,h^{-1}\,$Mpc (eq.~\ref{eq:cgamma(rp)}). With
the renormalization factors determined, we convert $w_{p}(r_{p})$ and
$w_{\pi}(\pi)$ into $\xir$ and $\xip$ using
eq.~[\ref{eq:xi(r_p)}--\ref{eq:x(pi)}].

As discussed above, $\xir$ should be close to the true 3D correlation
function $\xi(r)$, while $\xip$ should be distorted by peculiar
motions of the AGN host galaxies. Indeed, we observe a suppression in
$\xip$ relative to $\xir$ at separations
$\pi<1\,h^{-1}\,$Mpc. Unfortunately, the corresponding radial velocity
difference, $\Delta z\approx0.0003$, is uncomfortably close to the
uncertainties in the AGES redshift measurements (Kochanek et al., in
preparation). Therefore, we ignore the $\xip$ data at
$\pi<1\,h^{-1}\,$Mpc. At intermediate separations
($r\approx[1-10]\,h^{-1}\,$Mpc), $\xip$ is expected to be enhanced by
the peculiar motions (see below). No such enhancement is present in
the data. In fact, the ratio $\xip/\xir$ is fully consistent with 1 at
separations $>1\,h^{-1}\,$Mpc within the measurement
uncertainties\footnote{The uncertainties of the $\xip/\xir$ ratio are
  computed by propagation of uncertainties in $\xip$ and $\xir$. In
  principle, a region in the 2-dimentional $\xi(r_{p},\pi)$ enters
  into both projections. However, the analysis shows that the data in
  this region contribute negligibly to the ``signal'' and ``noise'' in
  the ratio, so our simple uncertainty calculation is adequate.}
(Fig.~\ref{fig:ksi}b). Below, we show that this can be used to put an
upper limit on the fraction of AGNs that can reside in satellite
galaxies orbiting within massive dark matter halos.

\subsubsection{Comparison with previous measurements}

Our results represent the most accurate measurement of the spatial
clustering of X-ray selected AGNs to date, so a comparison with
earlier observations is useful.

The first detection of angular clustering of X-ray sources was
reported by \cite{1995ApJ...455L.109V} based on the analysis of the
\emph{ROSAT} PSPC data. Using the Limber equation reconstruction
\citep{peebles80}, these authors estimated a correlation length of
$19\pm5\,h^{-1}$\,Mpc (comoving) at $z=1.5$ using raw measurements,
and $7.5\pm2\,h^{-1}$\,Mpc after correcting for the ``amplification''
bias caused by the poor angular resolution of the \emph{ROSAT}
PSPC. The latter value is in good agreement with our results.

Our results also are in good agreement with direct measurements of the
spatial clustering in the ROSAT NEP survey
\citep{2007A&A...465...35C}, \emph{Chandra} surveys in Deep Fields
North and South \citep{2005A&A...430..811G}, and the \emph{XMM-Newton}
COSMOS field \citep{2009A&A...494...33G}. The correlation length we
measure in the low redshift bins (see \S\ref{sec:res:vmin-z} below) is
in good agreement with $r_{0}=5.5-6\,h^{-1}\,$Mpc found in two studies
based on cross-correlations of X-ray AGNs with galaxy catalogs
\cite{2009ApJ...696..891H,Coil:2009hg}, although is higher than the
$r_{0}=4.3\pm0.35\,h^{-1}\,$Mpc reported in a similar work by
\cite{Krumpe:2010hz}.

Correlation lengths for the X-ray AGNs have been estimated also from
the Limber inversions of the angular clustering measured for the
\emph{XMM-Newton} and \emph{Chandra} sources. A wide range of $r_{0}$
values can be found in the literature \citep[e.g.,][and references
therein]{2008ApJ...674L...5P}; our measurements are inconsistent with
$r_{0}>10\,h^{-1}$~Mpc reported in some of these analyses.

\section{AGN Clustering Model}
\label{sec:HOD}

In the past ten years, a very successful framework for modeling the
nonlinear clustering properties of galaxies has been developed
\citep[the so called Halo Occupation Distribution (HOD) model,
see][among
others]{2000MNRAS.318..203S,2000ApJ...543..503M,2000MNRAS.318.1144P,2001ApJ...550L.129W,2001ApJ...546...20S,2002ApJ...575..587B,2003ApJ...593....1B}. The
approach is based on the idea that the distribution of dark matter can
be fully described through the mass function, linear bias, and density
profiles of dark matter halos. These elements are well-calibrated
using $N$-body simulations. The two additional ingredients of the
model, which are less well known, are the probability distribution for
a halo of mass $M$ to contain $N$ galaxies, and the distribution of
galaxies within the halos. These functions can be parameterized by the
functions suggested by the results of high-resolution numerical
simulations \citep[e.g.,][]{2004ApJ...609...35K}, and some parameters
of the model can be in fact determined by fitting the observed
correlation functions. In particular, \cite{2004ApJ...609...35K} and
\cite{2005ApJ...633..791Z} show that the elements of HOD can be
effectively decomposed into two components, separately describing the
properties of central and satellite galaxies within the dark matter
halos.

The HOD model is now very well developed for fitting the \emph{shape}
of the two-point correlation function. This technique has been applied
to modeling the projected two-point correlation functions, $w_{p}(r_{p})$,
for objects ranging from Lyman-break galaxies at $z=3-5$
\citep{2006ApJ...647..201C} to relatively low-$z$ quasars from SDSS
\citep{2009MNRAS.397.1862P}. Recently, the HOD approach has been
developed also for modeling the redshift-space distortions in the
galaxy correlation functions
\citep{2006MNRAS.368...85T,2007MNRAS.374..477T} --- just the type of
information we are aiming to use in this work to constrain the
locations of Bootes/\emph{Chandra} AGNs within the host dark matter
halos.

In principle, the HOD models for galaxy clustering are analytic, and
thus are convenient for those applications in which the cosmological
parameters are varied. However, some of the most essential parameters
of the HOD models are calibrated using numerical simulations. If one
is interested in varying the parameters of galaxy distribution at a
fixed redshift in a fixed cosmological model, it is more accurate ---
and easier --- to obtain the model correlation functions directly from
numerical simulations rather than to rely on analytic approximations
derived from analyzing the simulations. This is the approach we take
here.

\subsection{Numerical Simulations}


The set of simulations we use in this work is described in
\cite{2004ApJ...614..533T} and \cite{2006ApJ...647..201C}. These are
high-resolution dissipationless simulations run in a flat $\Lambda$CDM
cosmology with parameters close to the present-day ``concordance''
values, $\Omega_{\rm M}=0.3$, $h=0.7$, and $\sigma_{8}=0.9$. The
simulations follow the evolution of dark matter in a
$120\,h^{-1}\,$Mpc box. The box contained $512^{3}$ dark matter
particles with mass $m_{p}=1.07\times10^{9}\,h^{-1}\,M_{\odot}$; the peak
resolution reached $1.8\,h^{-1}\,$kpc.

The locations and velocities of the dark matter particles in the
simulations were recorded at $z=0.09$, 0.5, 1.0, 2.0, 3.1, 4.0, and
5.0, matching well the redshift distribution in the
Bootes/\emph{Chandra} AGN sample (see Fig.\ref{fig:dndz}). The
simulation outputs were then analyzed to identify dark matter halos
and subhalos using a modification of the \cite{1999ApJ...516..530K}
bound density maxima halo-finding algorithm \citep[see][for
details]{2004ApJ...609...35K}. The main steps of this procedure are
identification of local density peaks, and analysis of the density,
circular velocity, and velocity dispersion profiles with simultaneous
removal of unbound particles. The final profiles, using only bound
particles, are used to calculate the halo properties such as the
circular velocity profile $V_{\rm circ}=[GM(<r)/r]^{1/2}$ and the
maximum circular velocity, $\vmax$. The completeness limit for halo
identification using this procedure is $\sim 50$ particles. The
corresponding $\vmax$ limit is $\sim 80\,$\kms, and the associated
mass limit is $\sim 5\times10^{10}\,M_{\odot}$. The best-fit power law
to the $\vmax-\M200$ relation\footnote{Hereafter, we use
  the mass defined within an overdensity threshold of 180 with respect
  to the mean density of the Universe at the given
  redshift. The correspondence between $\vmax$ and mass is
quoted for $z=1$ unless the redshift is stated explicitly.}  derived in these simulations at $z=1$ is
\begin{equation}
  \log \M200 = 4.57 + 3.21\,\log\vmax,\label{eq:M-vmax}
\end{equation}
where the velocities are in units of \kms{} and masses are in units of
$h^{-1}\,\Msun$. 

The identified halos were then classified into host halos whose
centers are not located within any larger virialized systems, and
subhalos which lie within the virial radius of a larger system
\citep[see][for details of this
procedure]{2004ApJ...614..533T}. Briefly, a halo is classified as a
subhalo if its center is within $r_{180}$ of the center of a more
massive halo, where $r_{180}$ is the radius which corresponds to a
mean spherical overdensity of 180 relative to the mean density at the
given redshift. In the real Universe, centers of host halos can be
identified as locations of the groups' central galaxies, while
subhalos correspond to satellite galaxies.

\begin{figure*}
\centering
\includegraphics[width=1.999\columnwidth]{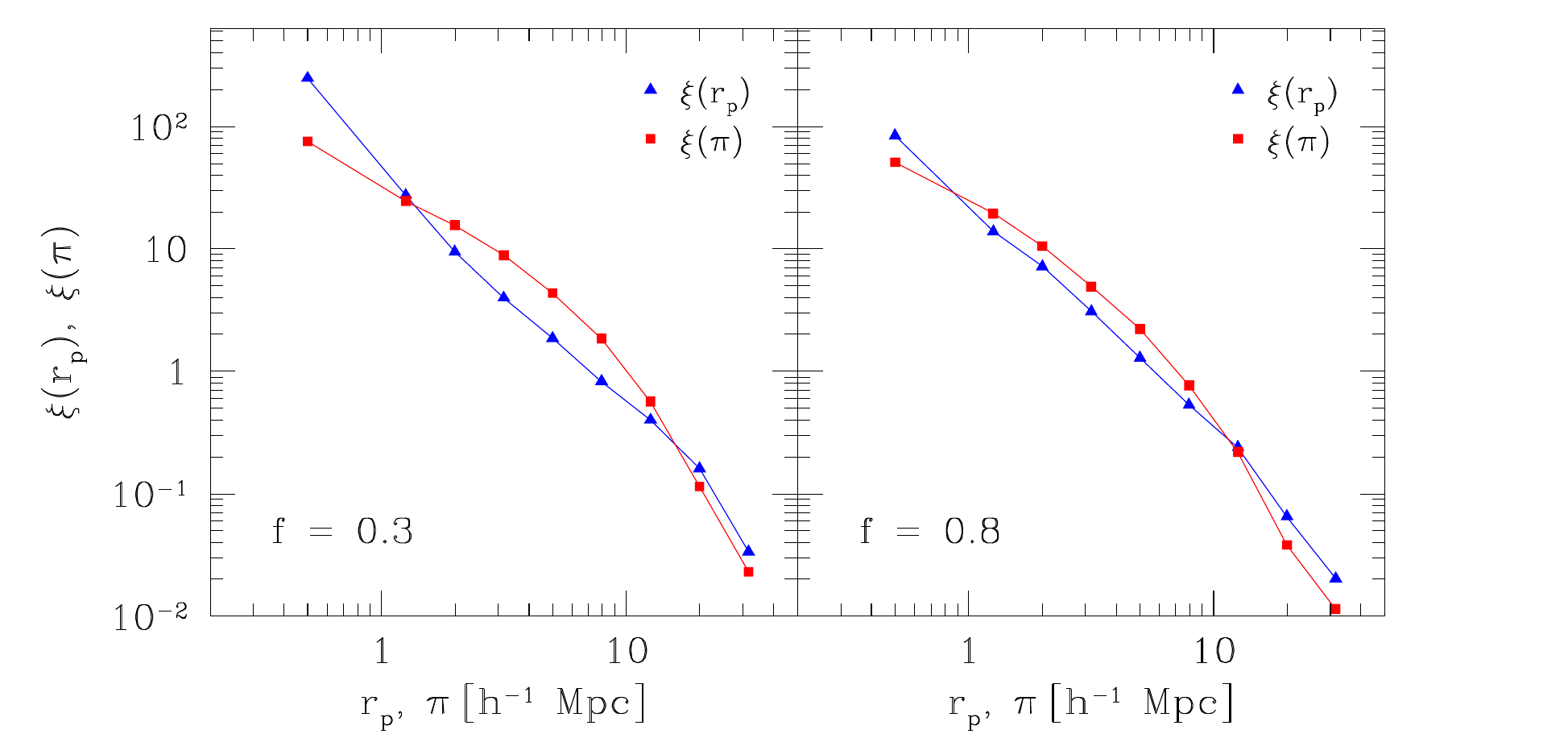}
\caption{The projected correlation functions for halo samples at $z=1$
  with $\vmax>250 \,\,{\rm km \,\, s^{-1}}$, calculated with
  $\fcen=0.3$ and $\fcen=0.8$. The correlation functions for smaller
  $\fcen$ (more AGNs in the satellite galaxies) show a stronger
  ``1-halo'' term at small separations [$\xir$] and a stronger
  presence of the ``finger of God'' effect [$\xip$].}
\label{fig:plot_f}
\end{figure*}

\begin{figure*}
\centering
\includegraphics[width=1.999\columnwidth]{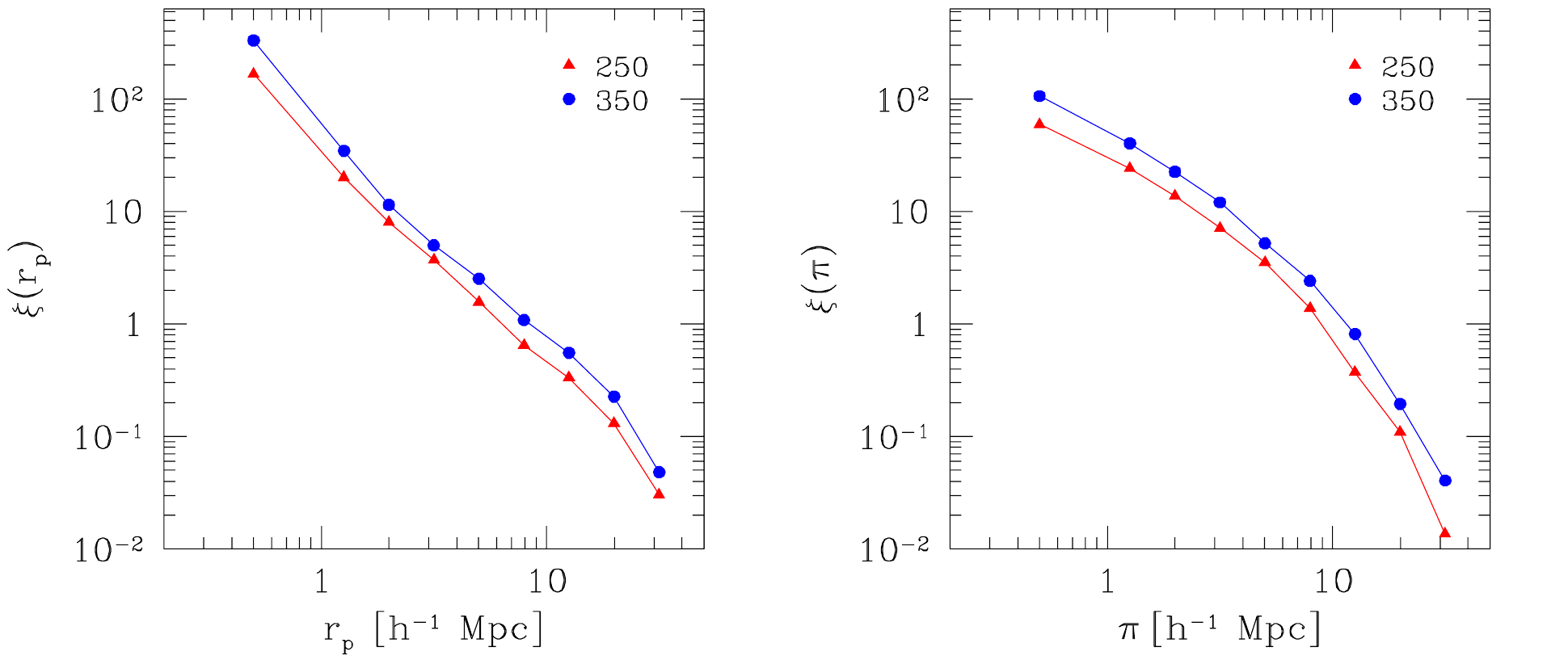}
\caption{The projected correlation functions of halo samples at $z=1$,
  calculated for $\Vmin=250$ and~350\,\kms{} and $\fcen=0.5$. The trend
  with $\Vmin$ is mainly equivalent to a uniform scaling of the
  correlation amplitude at all separations.}
\label{fig:vel}
\end{figure*}

\subsection{Model of the AGN population}

Our data can constrain the following two basic properties of the AGN
population. First, the overall clustering amplitude observed in $\xir$
constrains the mass scale of the AGN host dark matter halos. Note that
it is the mass scale of host halos, not of individual subhalos in
which the AGNs might reside, which is constrained by the $\xir$
amplitude. For example, a population of small subhalos with
$\vmax=50\,$\kms{} which are located within larger halos with
$\vmax=300\,$\kms{} has the clustering length which closely
matches that for larger, parent host halos, and not that for the
entire population of $\vmax=50\,$\kms{} halos. Therefore, the first
parameter of the model we use to characterize the spatial distribution
of AGNs is the minimum $\vmax$ for the halos which can contain the
X-ray AGNs. The AGN can be located either at the center of such a halo
or in any of its smaller subhalos.

Note that we choose to characterize the halos using their $\vmax$
rather than the virial mass for the reasons outlined in
\cite{1999ApJ...520..437K}, \cite{2005ApJ...618..557N}, and
\cite{2006ApJ...647..201C}. The maximum circular velocity is a more
direct measure of the depth of the halo potential wells. It reflects the
central properties of the halo better, and is less subject to the
effects of tidal stripping, than the halo virial mass which is
dominated by the matter at large radii. Therefore, we can expect that
the stellar content of the halo and all baryonic processes in the
center, including the AGN activity, are better correlated with $\vmax$
than with $M_{\rm vir}$. \cite{2005ApJ...618..557N} and
\cite{2006ApJ...647..201C} argue further that the best indicator for
the stellar mass of \emph{subhalos} is their $\vmax$ before accretion
onto the host halo. Since our results are not very sensitive to the
$\vmax$ threshold for subhalos (see below), we do not make this
distinction.
%
%

For a fixed threshold of the host halo $\vmax$, the correlation length
of astronomical objects only weakly depends on whether these objects
are located in the halo central galaxies or in smaller subhalos. As
discussed above, the ratio $\xip/\xir$ should be a much more sensitive
and direct indicator for the satellite fraction. We parameterize this
fraction by the probability, $\fcen$, for objects to reside in the
central galaxies of host halos; $1-\fcen$ is, therefore, the
probability for objects to reside in any of the smaller subhalos. The
two parameters, the $\vmax$ threshold, $\Vmin$, and the probability
$\fcen$, fully specify the relation between our model AGN
population
and the dark matter halos and subhalos identified in the numerical
simulations. $\Vmin$ is primarily constrained by the correlation
length of $\xir$, and $\fcen$ is mostly constrained by the $\xip/\xir$
ratio\footnote{Note that we assume that $\fcen$ is independent of
$\vmax$ and we have not explored the models in which $\fcen$ is very
different for the most massive halos. Effectively, our derived constraints
correspond to a mean $\fcen$ in the velocity range
$\Vmin<\vmax\lesssim2\Vmin$, containing $\sim 90\%$ of halos with
$\vmax>\Vmin$ \citep{Klypin:2010p3161}.}.

Algorithmically, we simulate the AGN locations by randomly drawing the
halos and subhalos from the simulation box according to the parameters
$\Vmin$ and $\fcen$. First, we select all bound structures (both halos
and subhalos) with $\vmax>80\,$\kms. This threshold is slightly higher
than the resolution limit in the simulations. We verified that the
final results are nearly the same when this initial threshold is
varied between 50\,\kms{} and~100\,\kms. We then select only those
subhalos which are contained within halos with $\vmax$ above the given
value of $\Vmin$.  We then randomly put a small number of objects
(10--100) within the simulation box\footnote{This is done to
  approximate the low space density of \ChBootes{} AGNS.}.  With the
probability $1-\fcen$, the object is associated with one of the
selected subhalos, and with the probability $\fcen$ it is put in the
center of one of the halos. This procedure is repeated multiple times
randomly selecting one of the box axes as the line of sight. Using
these simulated objects, we derive a model of $\xir$ and $\xip$ for
each combination of $\Vmin$ and $\fcen$.


\subsection{Model Correlation Functions $\xir$ and $\xip$}
\label{sec:model:corr:functions}

The correlation functions $\xir$ and $\xip$ were derived from the
simulation outputs on a grid of parameters within the range
$\Vmin\in[200;370]\,$\kms{} and $\fcen\in[0;1]$. Examples are shown in
Fig.~\ref{fig:plot_f} and~\ref{fig:vel}. Comparison of the two panels
in Fig.~\ref{fig:plot_f} illustrates the effect of $\fcen$ on the
correlation functions. For smaller $\fcen$ (more objects located in
satellite galaxies, left panel), the projected correlation function
$\xir$ shows a stronger component at $d<1\,h^{-1}\,$Mpc in excess of a
power law extrapolation from larger separations. This excess is
attributed to the ``1-halo'' term in the analytic halo model of the
correlation function. At the same time, $\xip$ shows a larger
suppression of the correlation amplitude at $d<1\,h^{-1}\,$Mpc with
respect to $\xir$ at the same separations, and a stronger enhancement
over $\xir$ at $d=2-10\,h^{-1}\,$Mpc for small $\fcen$. This is the
consequence of a stronger ``finger of God'' effect in the case when
more objects are located in the satellite galaxies. Unfortunately, the
statistical uncertainties in the real data do not allow a detailed
modeling of the observed $\xir$ at small separations. Modeling of the
$\xip$ at small separations is further complicated by the effect of
uncertainties in the redshift measurements
(\S~\ref{ssec:fun:results}). At large separations, $\gtrsim 0.1$ of
the simulation box size, the correlation functions derived from the
simulations are not reliable
\citep{2004ApJ...609...35K,2006ApJ...644..687C,2006ApJ...647..201C}. Taking
all these considerations into account, we will match the observed and
model correlation functions in the intermediate range of radii,
$d=1-12\,h^{-1}\,$Mpc.

\begin{figure}
\centerline{
\includegraphics[width=0.99\columnwidth]{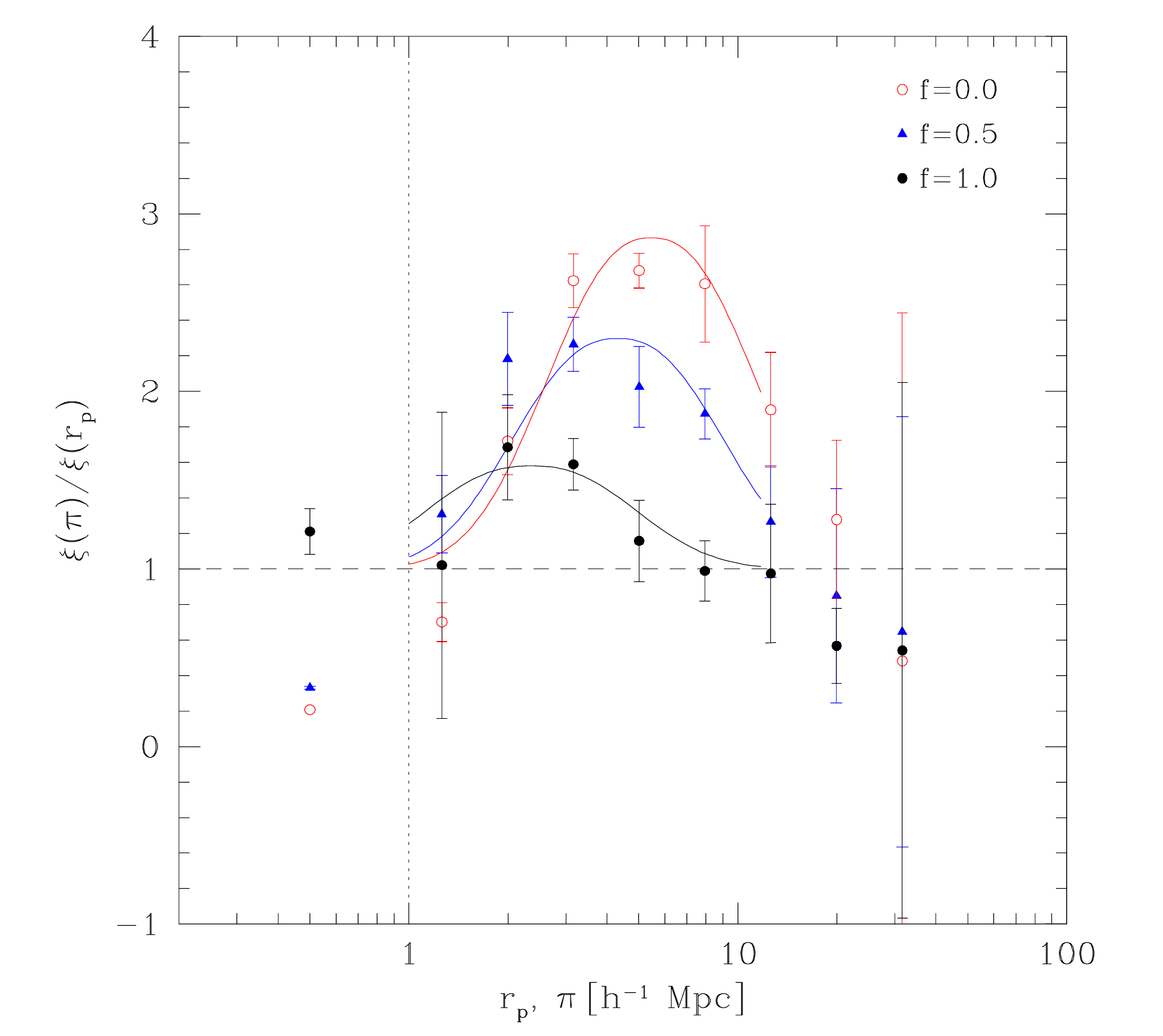}}
\caption{The ratio of the model correlation functions computed for
  $\fcen=0$, $0.5$, $1$, and $\Vmin=310\,$\kms. Uncertainties at each
  separation are estimated using the variance of the ratio computed
  for 3 different viewing angles and reflect mostly the cosmic
  variance within the simulation box. The solid lines shows the
  analytic approximation (eq.~\ref{frm:gauss}). Note that the analytic
  fits are derived from the global model with smoothed trends of
  parameters as functions of $f$ and $\Vmin$ (see text). The models
  are shown in the range $1-12\,h^{-1}\,$Mpc which we use in in
  applying these models to the data.}
\label{fig:ratio:a}
\end{figure}

First, we compute the correlation length, $r_{0}$, for each
combination $(\Vmin,\fcen)$. This is done by fitting a power law
function, $(r_{p}/r_{0})^{-\gamma}$, to $\xir$ in the range
$r_{p}=1-12\,h^{-1}\,$Mpc.  We then compute the ratio $\xip/\xir$ (an
example is shown in Fig.~\ref{fig:ratio:a}) and fit it in the same
range of separations with a modified log-normal function,
\begin{equation}\label{frm:gauss}
\phi(d)=1+A\exp\left(-\left[\frac{(\log d -\log d_{0})^2}{2D^2}\right]^{g}\right).
\end{equation}
The index $g$ is fixed at the mean best-fit value for all
$(\Vmin,\fcen)$ combinations, $g=1.2$. The $\xip/\xir$ ratio derived
from the simulations shows substantial variations related to cosmic
variance (this can be estimated by comparing the results for three
viewing angles, see errorbars in Fig.~\ref{fig:ratio:a}). Therefore,
we need to smooth the results of the fit by eq.~\ref{frm:gauss}. This
is achieved by fitting low-order polynomials to the parameters $D$,
$d_{0}$, and $A$ as a function of $\Vmin$ and $\fcen$. We found that
an adequate description is achieved if the best-fit values of $D$ and
$d_{0}$ are approximated as a linear function of $\fcen$, and
$A(\Vmin,\fcen)$ is fit by a second-order polynomial. An example of
the fitting function derived from this smooth map is shown in
Fig.~\ref{fig:ratio:a}. Due to the size of the simulation box, the
uncertainties in the smoothed model are still finite, but we verified
that they are negligible compared to those in the data.

\subsection{Application to the Data}

In applying the correlation function model to the data we avoid
including any sensitivity to the \emph{slope} of the correlation
function. The primary motivation is that our method assigning the AGN
locations to the dark matter halos may be overly simplistic to
correctly predict the details of the correlation function
\emph{shape}. Also, the cosmological parameters used in the simulation
are slightly different from the currently accepted values, resulting
in a systematic difference in the shape of the perturbation power
spectrum in the simulated and real universes. This said, the models
derived from simulations do provide a good fit to the $\xir$ data (see
discussion in \S\,\ref{sec:res:vmin-f} below).

Based on these considerations, our $\chi^{2}$ includes two components.
First, we use the value of the correlation length derived from fitting
the $\xir$ function (\S~\ref{ssec:fun:results}). Second, we use the
ratio $x=\xip/\xir$ in the range of separations $1-12\,h^{-1}\,$Mpc
(the data at $\pi<1\,h^{-1}\,$Mpc are not used because they are likely
affected by the redshift measurement uncertainties, see
\S\ref{ssec:fun:results}). For halos with circular velocities
$\vmax\sim 300\,$\kms{} (as indicated by the amplitude of the AGN
correlation function, see below), the ``fingers of God'' extend to
$\vmax/H\sim 3\,h^{-1}\,$Mpc, just in the middle of this range of
separations. Formally, the constraints on the parameters of the AGN
population model, $\Vmin$ and $\fcen$, are derived using a $\chi^{2}$
function computed as
\begin{equation}
  \label{eq:chi2}
  \chi^{2} = \frac{(r_{0}-r_{0}^{\rm mod})^{2}}{\sigma_{r0}^{2}} +
  \sum \frac{(x-x^{\rm mod})^{2}}{\sigma_{x}^{2}}
\end{equation}
where the summation in the second term is over the data points in the
$1-10\,h^{-1}\,$Mpc separation range, and the model functions are those
described in \S~\ref{sec:model:corr:functions}.

\begin{figure*}
\includegraphics[height=0.8\columnwidth]{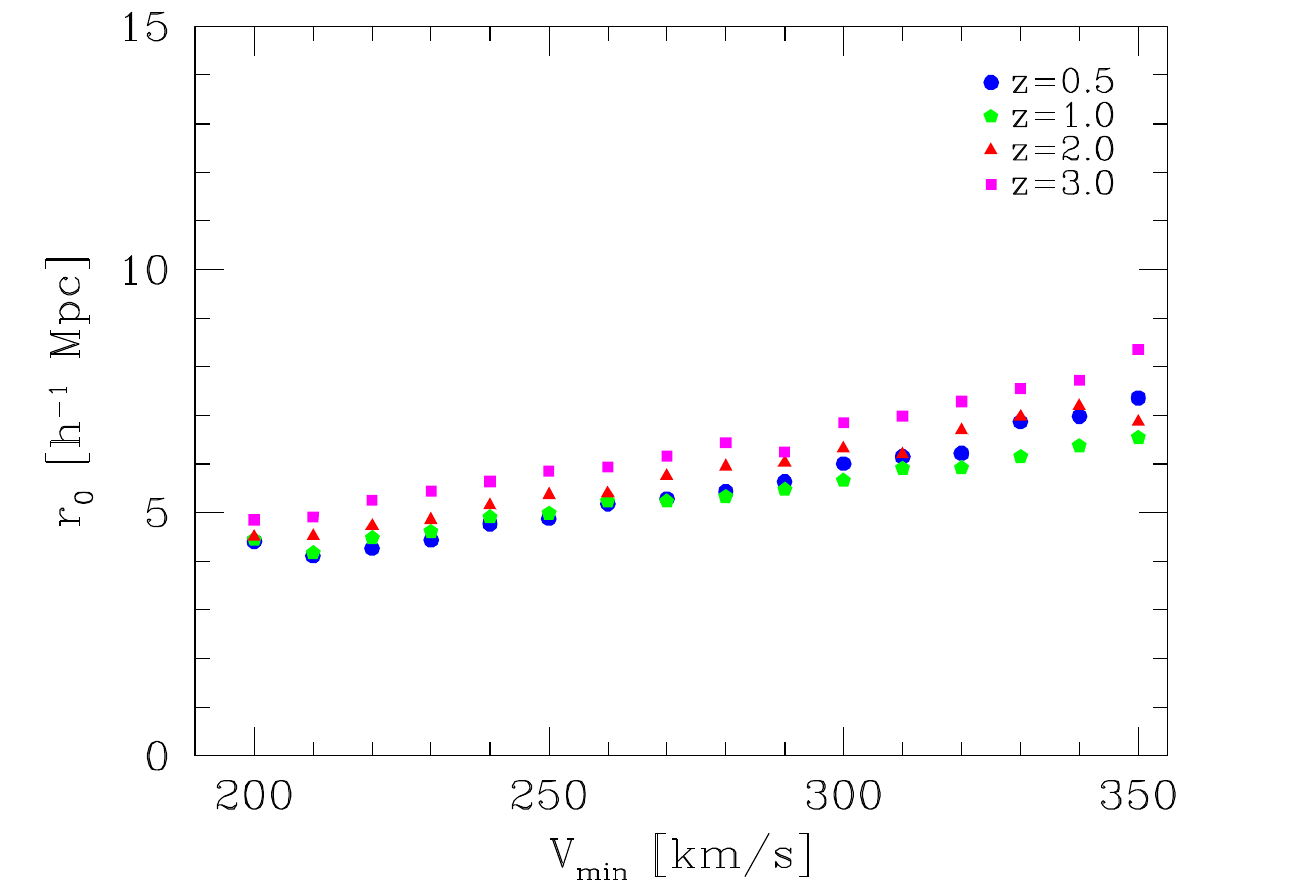}\hfill
\includegraphics[height=0.8\columnwidth]{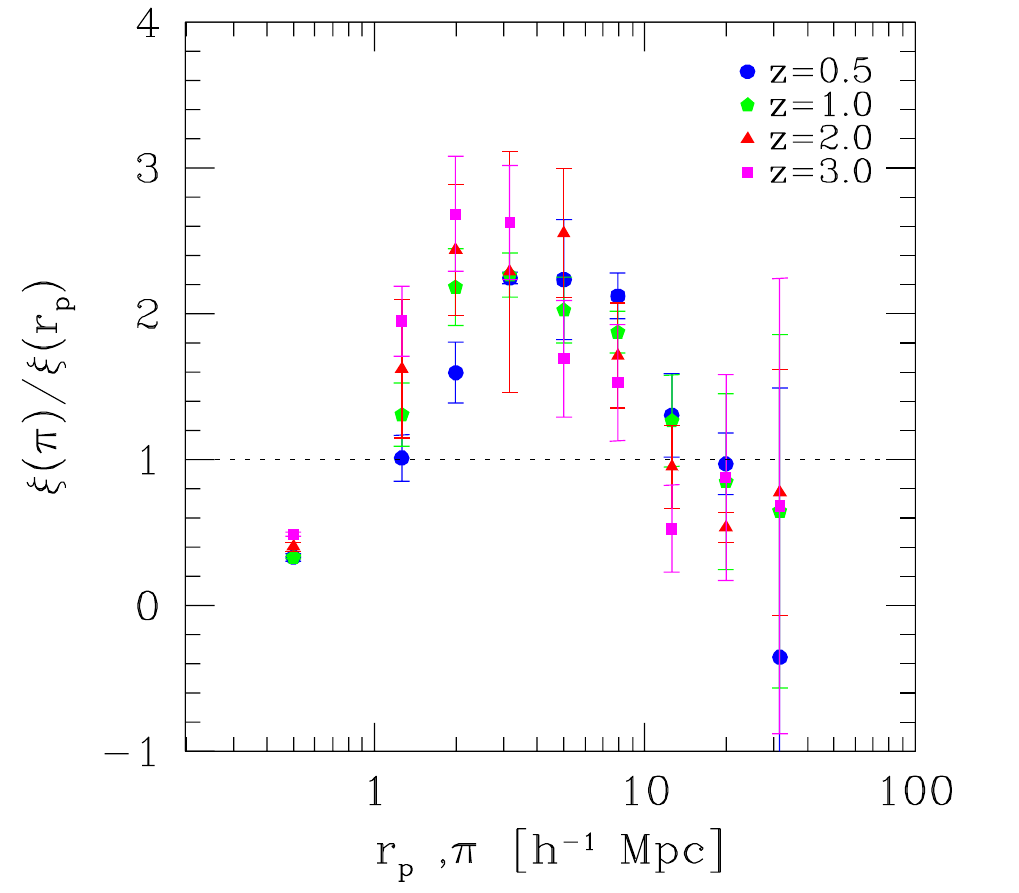}
\vspace*{-2mm}
\caption{\emph{Left:} Dependence of $r_0$ on $\Vmin$ for the objects
  at the centers of distinct halos ($\fcen=1$) for the simulation
  outputs at $z=0.5$, 1, 2, and 3. There is almost no change with
  redshift for $z\lesssim2.5$. \emph{Right:} the ratio of $\xip/\xir$
  for the population model with $\Vmin=310\,$\kms{} and $\fcen=0.5$
  (50\% of objects are in the satellite galaxies of the
  $\vmax>310\,$\kms{} halos). Any difference in this ratio between
  different simulation snapshots is within the uncertainties
  (estimated from analyzing 3 different projections for each
  snapshot).}
\label{fig:z-trend}
\end{figure*}

\subsubsection{Redshift Trends}

The procedure described above provides a correlation function model at
the redshifts where the simulation outputs were saved. The \ChBootes{}
AGNs span a wide redshift range and we have to bin the data into
several redshift intervals to achieve an acceptable level of accuracy
for the correlation function measurements. Therefore, we need to
account for any $z$-dependent trends of the correlation function
models.

Fortunately, for our choice of observables, $r_{0}$ and the
$\xip/\xir$ ratio, the redshift trends are very weak. This is
illustrated in Fig.~\ref{fig:z-trend}. The left panel shows $r_{0}$ as
a function of $\Vmin$ for a population model with $\fcen=1$ (all
objects are at the centers of distinct halos) for the simulation
outputs at $z=0.5$, 1, 2, and 3. Obviously, there is almost no change
in $r_{0}$ for a fixed $\Vmin$ at $z<3$. Any changes are much smaller
than the uncertainties of our $r_{0}$ measurement even for the full
sample. Therefore, we conclude that the model $r_{0}$ as a function of
$\Vmin$ does not evolve over our redshift range of
interest.\footnote{Note that $r_{0}$ as a function of \emph{mass} does
  evolve with redshift, as expected. However, this evolution appears
  to be canceled by the evolution in the $M-\vmax$ relation and the
  trend of $r_{0}$ with $M$ at a given redshift.}  The ratio
$\xip/\xir$ also shows little, if any, evolution with redshift. The
right panel in Fig.~\ref{fig:z-trend} shows the results for the
population model with $\Vmin=310\,$\kms{} and $\fcen=0.5$ (50\% of
objects are in the satellite subhalos of $\vmax>310\,$\kms{}
halos). Any difference between the simulation outputs is within the
uncertainties (estimated from analyzing three different projections
for each simulation output).

The lack of evolution in the clustering model over our redshift
interval (and also the lack of detectable evolution of $r_{0}$ with
$z$, see \S\,\ref{sec:res:vmin-z} below) indicates that we can safely
combine the data over the entire redshift range in the
sample. Furthermore, there is no need to weight the models with the
redshift distribution --- one simply can use the results for the
simulation output at $z=1$. We take this approach in fitting the
parameters of the population model, $\Vmin$ and $\fcen$
(\S~\ref{sec:res:vmin-f}). In addition, we constrain the evolution of
$\Vmin$ with $z$ (\S~\ref{sec:res:vmin-z}) under the assumption that the
fraction of AGNs in the subhalos does not evolve (i.e., using a fixed
$\fcen$ derived from the analysis of the entire sample).

\subsubsection{Adjusting the Results to a Low-$\sigma_{8}$ Cosmology}
\label{sec:sigma8:corr}

Finally, we note that the numerical simulations we use were run
assuming a high value of the power spectrum normalization,
$\sigma_{8}=0.9$ at $z=0$. This results in an incorrect prediction of
the correlation amplitude for halos of a given mass, and thus slightly
biases the derived parameters of the AGN population model, in
particular, $\Vmin$. Obviously, it would be best to use the
simulations performed for the currently favored cosmological model
with $\sigma_{8}\approx0.8$ but in general, these were unavailable at
the time of this investigation. In
Appendix~\ref{sec:cosm-depend-scal}, we describe a procedure which can
be used to scale the results from the comparison with the simulation
to any desired cosmology. In particular, if we use the best-fit flat
$\Lambda$CDM cosmological model derived from the joint analysis of the
galaxy cluster mass function and other cosmological datasets,
$\sigma_{8}=0.786$ and $\Omega_{M}=0.268$ \citep{2009ApJ...692.1060V},
the masses reported below should be scaled by a factor of $0.69$, the
$\Vmin$'s decreased by 10\%, and the number density of objects
increased by 20\%.

\section{Modeling Results}

\subsection{$\Vmin$ and the Satellite Fraction for the Entire Sample}
\label{sec:res:vmin-f}

\begin{figure}
\centering
\includegraphics[width=0.99\columnwidth]{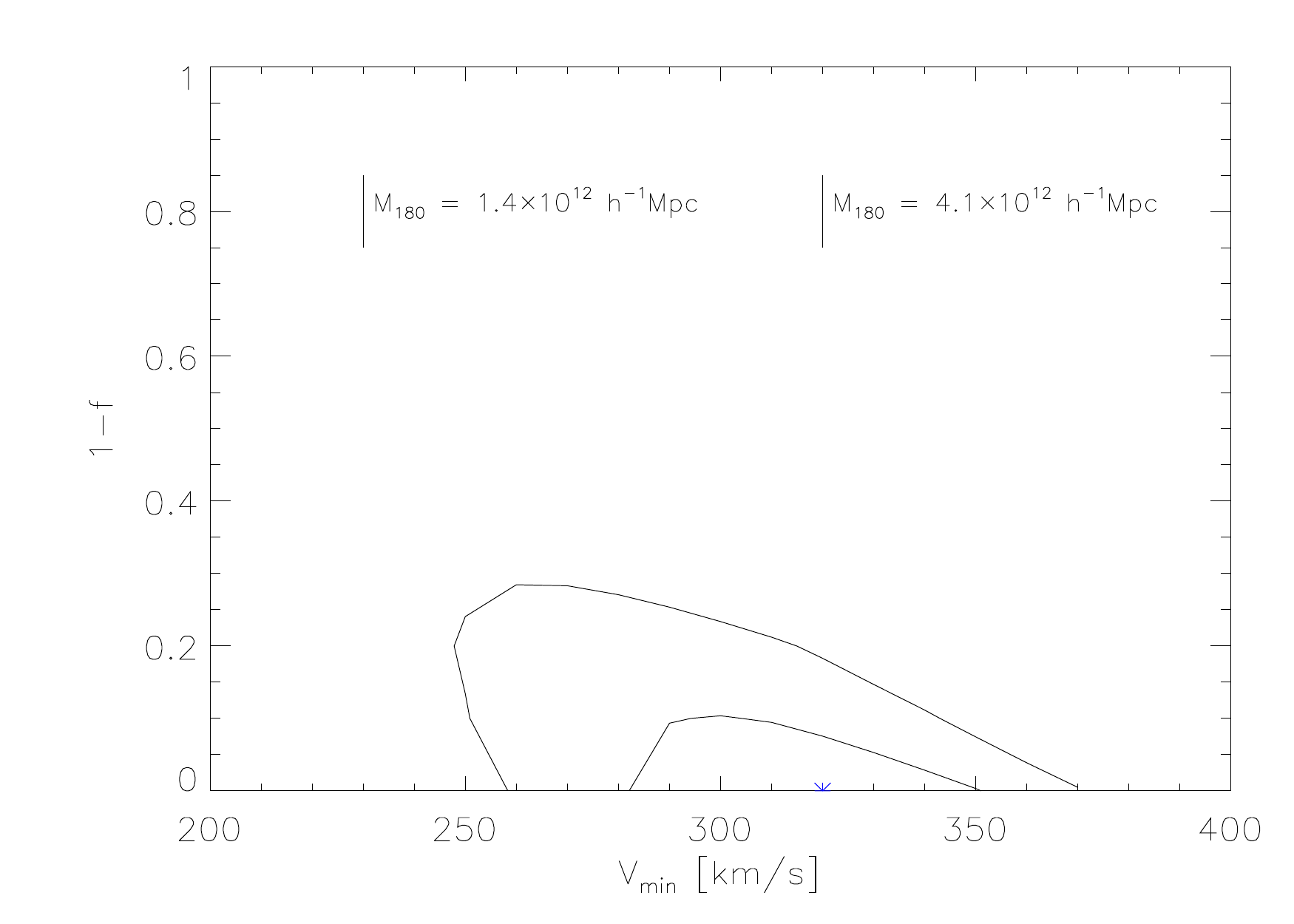}  
\caption{Combined constraints on halo occupation parameters, $\Vmin$
  and $\fcen$, obtained from the full sample of the \ChBootes{}
  AGNs. The best fit ($\Vmin=320\,$\kms{} and $\fcen=1$) is indicated
  with a star. The contours correspond to 68\% and 95\% confidence
  regions for two interesting parameters ($\Delta\chi^{2}=2.3$ and
  6.2, respectively). The quantity $1-f$ represents the fraction of
  AGNs residing in the satellite galaxies. We also indicate the dark
  halo masses corresponding to two values of $\Vmin$.}
\label{fig:Vmin-f}
\end{figure}

\begin{figure*}
\vspace*{-7mm}
\centerline{
\includegraphics[width=0.99\columnwidth]{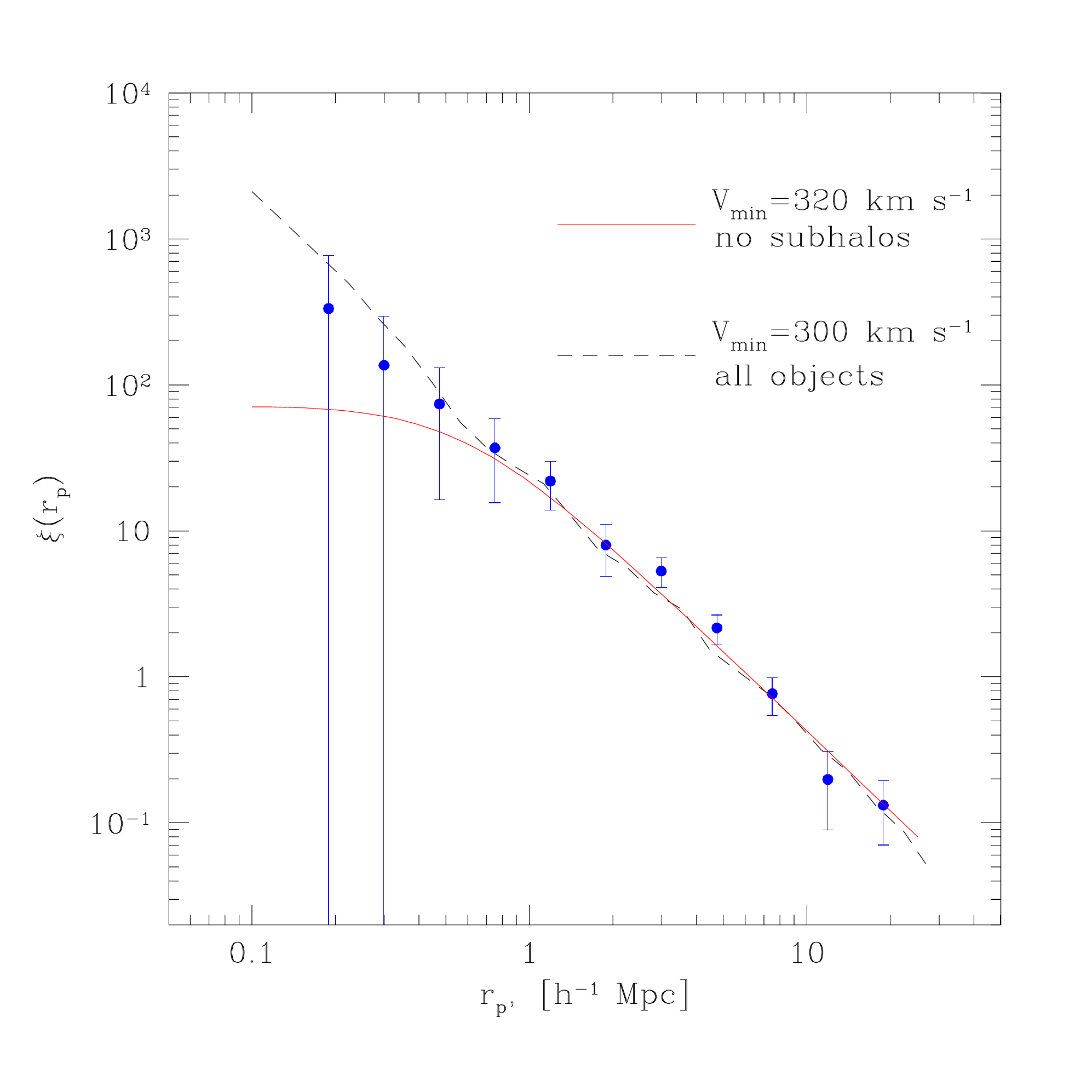}
\includegraphics[width=0.99\columnwidth]{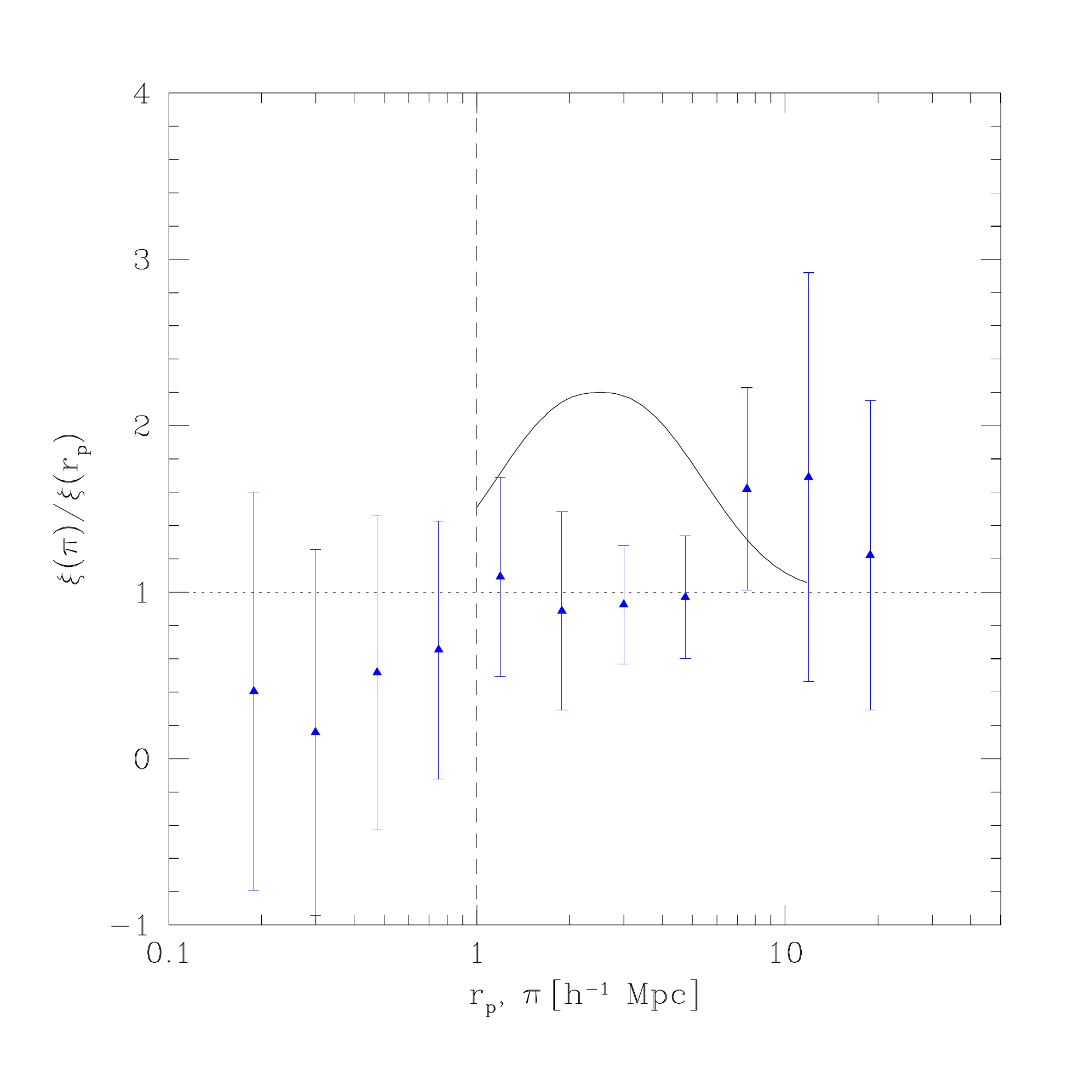}
}
\vspace*{-5mm}
\caption{\emph{(Left panel:)} Model correlation function for all objects in the
  simulation output (both halos and subhalos) with $\vmax>300\,$\kms
  (dashed line) in comparison with the observed $\xir$. Approximately
  11\% of objects with such $\vmax$ are satellites in larger halos,
  which results in the ``1-halo'' term visible as the excess in the
  correlation power at $r_{p}<0.7\,h^{-1}\,$Mpc relative to the model
  without subhalos (solid line). The peculiar motions of such halos
  lead to strong distortions of $\xip$ with respect to $\xir$ so that
  the ratio of $\xip/\xir$ reaches $\approx 2$ in the
  $2-6\,h^{-1}\,$Mpc separation range, inconsistent with the observed
  ratio \emph{(right panel)}}.
\label{fig:vmax300}
\end{figure*}

Figure~\ref{fig:Vmin-f} shows the combined constraints on the population
model parameters, $\Vmin$ and $\fcen$, obtained from fitting the full
sample of \ChBootes{} AGNs. The best-fit velocity threshold is
$\Vmin=320\pm20\,$\kms{} (68\% CL one-parameter uncertainty). At
$z=1$, this corresponds to
$\M200=4.1\times10^{12}\,h^{-1}\,M_{\odot}$, or
$2.4\times10^{12}\,h^{-1}\,M_{\odot}$ after correcting for the
lower-$\sigma_{8}$ cosmology (see \S~\ref{sec:sigma8:corr} and
Appendix~\ref{sec:cosm-depend-scal})\footnote{\cite{2009ApJ...696..891H}
  quote a higher mass, $\sim 10^{13}\,h^{-1}\,M_{\odot}$ for an X-ray
  AGNs sample with the mean $\langle z\rangle=0.51$, from which they
  measure $r_{0}=4.8\,h^{-1}\,$Mpc. The main source of the difference,
  as explained in \citet[see their p.1195]{2008ApJ...679.1192C}, is a
  more accurate model for the matter power spectrum at galactic scales
  deployed in modern simulations such as those we use here.}. We thus
conclude, in agreement with the earlier studies
\citep{2009A&A...494...33G,2009ApJ...696..891H}, that the X-ray AGNs at
high redshifts reside in small galaxy groups with masses of a factor
of a few above the present-day mass of the Milky Way.

As expected from the striking agreement of the observed $\xir$ and
$\xip$ projected correlation function, the best fit is $\fcen=1$,
i.e.\ \emph{all AGNs are located at the centers of distinct dark
  matter halos}. The 90\% CL upper limit on the satellite fraction is
$1-f<0.12$. This is significantly lower than the fraction of suitable
subhalos in the simulation box. For example, within the 1486 halos
with $\vmax>320\,$\kms{} there are 683 Milky Way-type subhalos
($\vmax=220\,$\kms), and many more smaller galaxies. If all these
galaxies had a uniform probablity to host an X-ray AGN, we would
expect $1-f>0.32$.

If we consider only the objects with $\vmax>320\,$\kms, 11\% of them
are satellites in more massive halos, which is comparable to the upper
limit for $1-f$ derived from our model. However, we still can exclude
a uniform probability for hosting an AGN in this case, because the
$\vmax>320\,$\kms{} subhalos by definition are members of more massive
halos, and thus have higher peculiar velocities than most of the
satellites in our $(\Vmin,f)$ model. To illustrate this point, we
computed the project correlation functions for all halos and subhalos
with $\vmax>300\,$\kms{} (Fig.~\ref{fig:vmax300}). This population has
nearly the same correlation length as our best-fit model with
$(\Vmin,f)=(320,1)$. In the $\xir$ function there is a strong
``1-halo'' component at $r_{p}\lesssim0.7\,$Mpc. We do not observe
such a component in the AGN autocorrelation function, even though the
quality of $\xir$ data at small scales is clearly insufficient for
distinguishing the models. However, the $\vmax>300\,$\kms{} dark
matter subhalos are mainly satellites in much more massive halos
(median $\vmax$ for their host halo is 530\,\kms). They have high
peculiar velocities and show a strong excess in $\xip$ over $\xir$ at
separations $1-10\,h^{-1}\,$Mpc, not present in the data.

Therefore, we can conclude that the X-ray AGNs at $z=1$ tend to
\emph{avoid} massive galaxies in the outskirts of yet more massive
groups and clusters, or satellite galaxies in the
$\vmax\gtrsim300\,$\kms{} galaxy groups. Instead, the AGNs are
preferentially located at the centers of distinct dark matter halos.

\subsubsection{Comparison with Previous AGN HOD Studies}

Our conclusion on the preferential location of AGNs in the halo
central galaxies contradicts the results of \citet[P09
hereafter]{2009MNRAS.397.1862P} who observe the presence of the 1-halo
term in the cross-correlation function of optically-selected $z<0.6$
quasars and Luminous Red Galaxies (LRGs), and use this to conclude
that a large fraction of the AGNs is hosted by satellite
galaxies. 

There are obvious differences in the AGN populations studied in P09
and our work. A striking mismatch in the derived host halo masses ---
$M>3\times10^{10}\,h^{-1}\,M_{\odot}$ estimated in P09 is a factor of
$\sim 10$ lower than our minimum mass --- indicates that we indeed may
be dealing with different object populations. However, we also note a
significant difference in the underlying methods. While our results
follow from the AGN \emph{auto}-correlation function, the P09 approach
is based on \emph{cross}-correlation of the AGNs with LRGs, the
tracers typically located in the halos a factor of $\sim 30$ more
massive than the AGN hosts (see, e.g., Table~A1 in P09).

Recently, \citet[M11 hereafter]{2011ApJ...726...83M} published the
halo occupation analysis for a sample of X-ray selected AGNs. Their
AGN sample is at lower redshifts ($z<0.36$) but closely matches ours
in terms of the estimated host halo mass. Similarly to P09, M11 detect
a 1-halo term in the cross-correlation function of AGNs and LRGs,
qualitatively indicating that some fraction of the AGNs may be located
in the satellite galaxies. However, even if the 1-halo term is
detected, the best-fit AGN HOD models of M11 imply that the AGN
incidence rate in the satellite galaxies decreases with increasing
halo mass, approximately in line with our conclusions above.

In summary, we can only speculate whether any differences in our
conclusions with those of P09 and M11 are due to not identical AGN
samples (optical vs.\ X-ray selection, or a substantial mismatch in
the typical X-ray luminosities in the M11 and our samples), or
different analysis methods (velocity space auto-correlation function
of AGNs vs.\ their spatial cross-correlation with more massive
tracers). A clear answer can be provided by detection of, or strong
upper limits on, the 1-halo term in AGN $\xir$ \emph{auto}-correlation
function. Unfortunately, the present data quality is insufficient for
this purpose.

\begin{deluxetable*}{p{2.5cm}cccccccc}
  \tablecaption{Clustering modeling results as a function of redshift\label{tab:values}
}
  \tablehead{
    \colhead{Redshift range} &
    \colhead{$z_{\rm med}$} &
    \colhead{$\langle\log L_{x}\rangle$} &
    \colhead{$r_{0}$} &
    \colhead{$\Vmin$} &
    \colhead{$\Mlim$} &
    \colhead{$n$}     &
    \colhead{$n_{\rm halo}$} &
    \colhead{$n/n_{\rm halo}$}
    \\[1mm]
    \colhead{}  &
    \colhead{(2)} & 
    \colhead{(3)} &
    \colhead{(4)} &
    \colhead{(5)} &
    \colhead{(6)} &
    \colhead{(7)} &
    \colhead{(8)} &
    \colhead{(9)}
  }
  \startdata
  $0.172-0.555$\dotfill & 0.374 & 42.58 & $5.52\pm0.56$ & $277\pm28$ & 
$3.7\pm1.1$ &
$1.54\times10^{-4}$&$1.32_{-0.30}^{+0.56}$ & $0.116\pm0.035$\\
  $0.555-1.000$\dotfill & 0.738 & 43.34 & $7.69\pm0.99$ & $383\pm35$ & 
$8.4\pm2.5$ &
$6.12\times10^{-5}$&$0.50_{-0.11}^{+0.21}$ & $0.123 \pm 0.037$\\
  $1.000-1.630$\dotfill & 1.279 & 43.85 & $6.89\pm1.11$ & $355\pm64$ & 
$4.9\pm2.7$ &
$2.56\times10^{-5}$&$0.49_{-0.17}^{+0.60}$ & $0.052\pm0.029$
  \enddata
\tablecomments{
Column (3) --- mean X-ray luminosity in the 0.5--2~keV band (rest-frame).
Column (4) --- best-fit correlation length assuming a fixed slope of
$\gamma=1.97$, in units of $h^{-1}$ comoving Mpc.
Column (5) --- threshold maximum circular velocity of the host dark
matter halos, in units of \kms{} (\S\ref{sec:res:vmin-z}).
Column (6) --- the halo virial mass corresponding to $\Vmin$, in units
of $10^{12}\,h^{-1}\,\Msun$.
Column (7) --- comoving number density of X-ray sources at
  $z_{\rm med}$, in units of $h^{3}\,$Mpc$^{-3}$.
Column (8) --- comoving number density of dark matter halos with $\vmax>\Vmin$ at
  $z_{\rm med}$, in units of $10^{-3}\,h^{3}\,$Mpc$^{-3}$.
Column (9) --- probability for the \ChBootes{} AGNs to be in active
state computed as the ratio of number density of AGNs and the dark
matter halos with $\vmax>\Vmin$.
}
\end{deluxetable*}

\subsection{$\Vmin$ as a Function of Redshift}
\label{sec:res:vmin-z}

\begin{figure}
\vspace*{-2mm}
\centerline{\includegraphics[width=0.999\columnwidth]{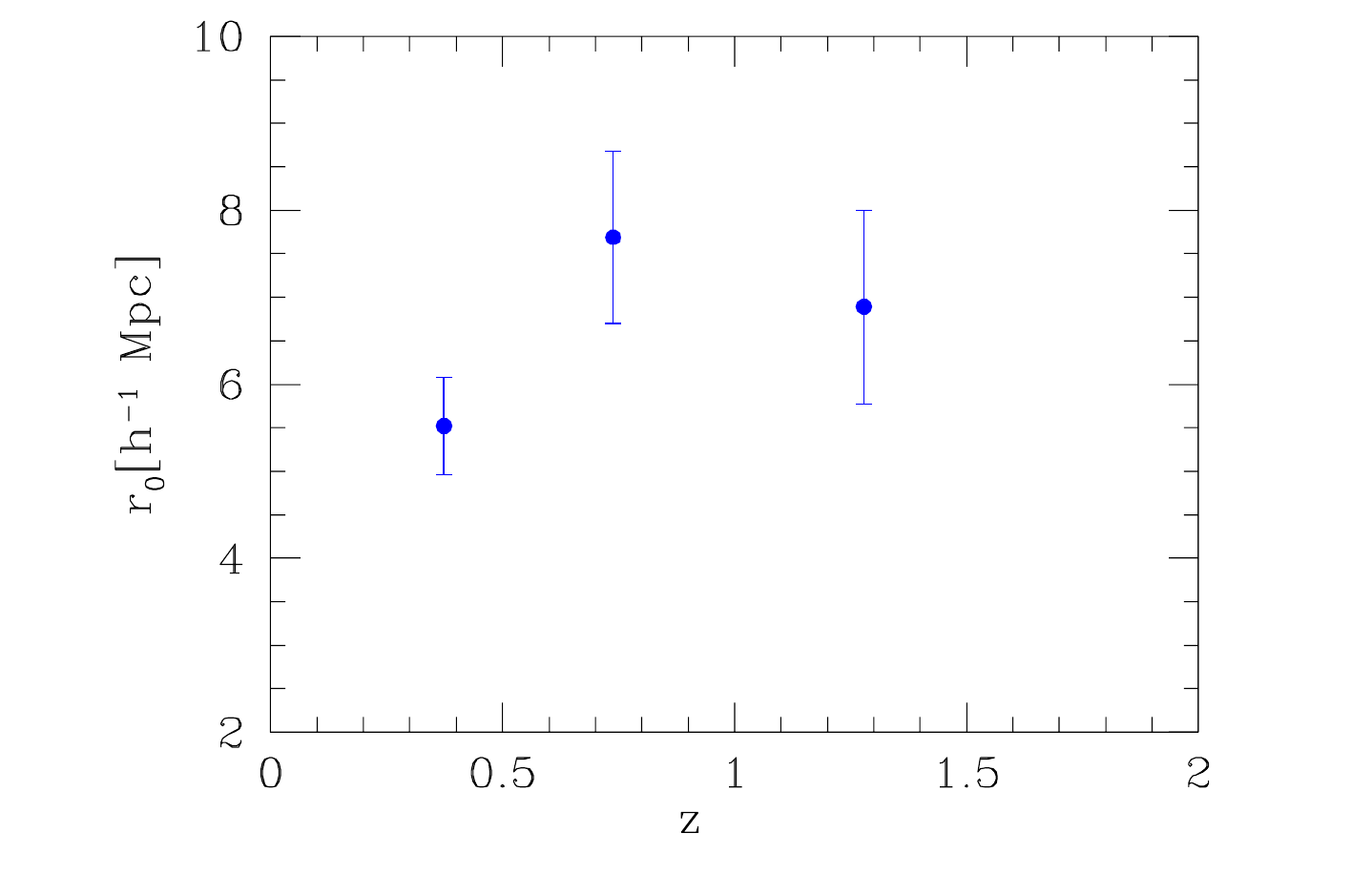}}
\caption{The correlation lengths, $r_{0}$, measured in 3 redshift
  intervals, $z=[0.172-0.555]$, $[0.555-1.000]$, and $[1.000-1.630]$,
  containing 320, 307, and 344 objects, respectively. The values of
  $r_{0}$ were obtained by fitting the projected correlation function,
  $w_{p}(r_{p})$, in each bin assuming the fixed slope of the correlation
  function $\gamma=1.84$ (the best-fit value for the entire sample).}
\label{fig:r0(z)}
\end{figure}

Splitting our \ChBootes{} AGN sample into 4 subsamples, $\sim 320$
objects in each, we can measure the correlation length, $r_{0}$, as a
function of redshift if we hold the slope of the correlation function
fixed at the best-fit value obtained for the entire sample,
$\gamma=1.84$. The $r_{0}$ in the highest redshift bin, $z=[1.63-5]$,
is poorly constrained, while in the other three samples we obtain
reasonably accurate values, given in Table~\ref{tab:values} and shown
in Fig.~\ref{fig:r0(z)}. There is almost no change in $r_{0}$ over the
redshift interval $z=0.17-1.6$, with all the measurements being
consistent with the average $r_{0}=6.41\,h^{-1}\,$Mpc derived for the
entire sample.

\begin{figure}
\centerline{\includegraphics[width=0.999\columnwidth]{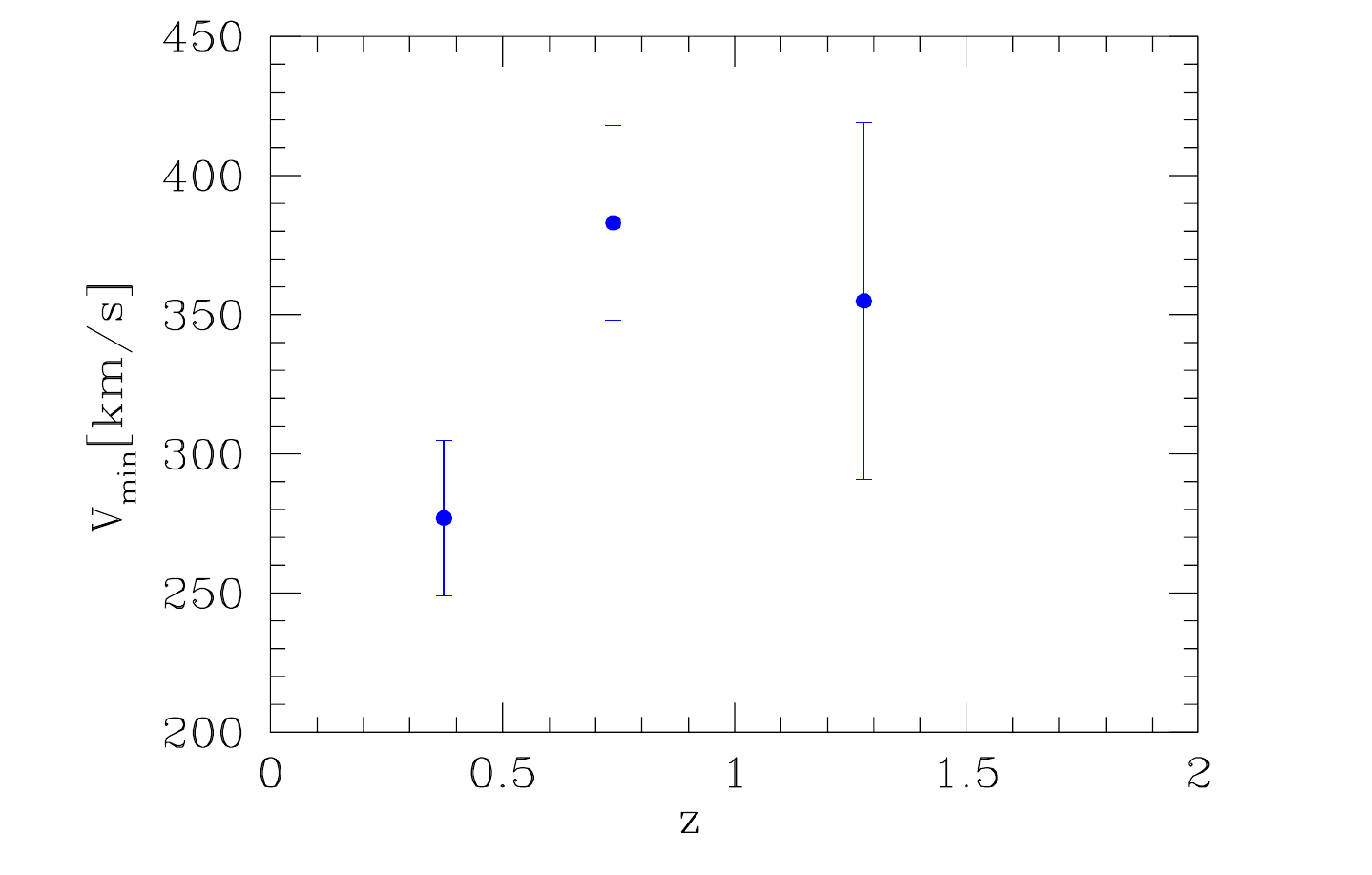}}
\caption{Minimum rotational velocity for dark matter halos,
  corresponding to the correlation lengths in Fig.~\ref{fig:r0(z)}.}
\label{fig:v(z)}
\end{figure}

Assuming further a fixed $\fcen=1$ at each $z$, as indicated by
modeling of the redshift-space distortions in the entire sample, we
can convert the best-fit values of $r_{0}$ at each redshift to the
threshold circular velocity for the parent dark matter halos. The
results are shown in Fig.~\ref{fig:v(z)}. We find no detectable trend
of $\Vmin$ or the corresponding mass threshold, $\Mlim$ (see
Table~\ref{tab:values}), with redshift, either. This appears somewhat
counterintuitive because in a flux-limited sample, such as ours, the
objects at higher redshift have higher intrinsic luminosities, and we
might expect them to be located in more massive dark matter halos. We
note, however, that the studies of optically selected QSOs also
indicate a weak or no trend of clustering length with the object
luminosity \citep{2005MNRAS.356..415C,2009ApJ...697.1656S}. Clustering
analysis of the SDSS quasars \citep{2009ApJ...697.1634R} shows mild or
no evolution of the real-space correlation length at $z\lesssim2$.

\subsection{Constraints on AGN Duty Cycle}

Having estimated the mass scale and therefore the space density of
\ChBootes{} AGNs, we can formally compute their duty cycle following
the approach of \cite{2001ApJ...547...12M}. Under the simplifying
assumption that the halo lifetime is approximately independent of
mass, the probability for the AGN to be active is simply
\begin{equation}\label{eq:duty:cycle}
  P_{\rm active}=\frac{n_{\rm AGN}}{n_{\rm halo}},
\end{equation}
where $n_{\rm AGN}$ is obtained from the $dN/dz$ fit
(eq.~\ref{frm:dndz}) for the median redshift for the given subsample.
The limiting X-ray luminosity is ill-defined for our sample because
the X-ray detections extend to very low limits in terms of the number
of detected X-ray photons \citep[$\ge4$, see][]{2005ApJS..161....9K},
and in this regime, a wide range of possible intrinsic intensities
corresponds to the given number of photons
\citep{2003ApJ...584.1016K}. However, as a guide for the typical
luminosity scale one can use the median $\log L_{x}$ reported in
Table~\ref{tab:values}. With these caveats, the probabilities given by
eq.~\ref{eq:duty:cycle} and reported in column~(9) of
Table~\ref{tab:values} correspond to the probability for the dark
matter halos with $\vmax>\Vmin$ to host an AGN with the instantaneous
soft-band X-ray luminosity of order $\langle\log L_{x}\rangle$ or
above. These $P_{\rm active}$ generally decline with mean luminosity
and/or redshift. Interestingly, our results indicate that AGNs are
quite common at low redshifts --- approximately $10\%$ of dark matter
halos with $\vmax>280\,$\kms{} (or mass
$\M200>3.7\times10^{12}\,h^{-1}\,M_{\odot}$) host an AGN with a
soft-band X-ray luminosity of $\sim 4\times10^{42}\,$\ergs{} or above.

\label{sec:duty:cycle}

\section{Discussion and conclusions}

We measured the clustering properties of X-ray selected AGNs using a
sample of 1282 sources with spectroscopic redshifts from the
9.3~deg$^2$ \emph{Chandra} survey in the Bo\"otes region --- one of
the most accurate such measurement to date. In agreement with previous
studies of X-ray and optically selected AGNs, we find that the
real-space correlation function can be approximated in the radial
range $1-20\,h^{-1}$ comoving Mpc by a power law with a slope of
$\gamma=1.84\pm0.12$. The correlation length is $\approx
6.4\,h^{-1}$\,Mpc, showing only weak trends with redshift at
$z=0.17-1.6$ (or with X-ray luminosity).

Matching the observed clustering properties of the \ChBootes{} AGNs to
those of dark matter halos in the high-resolution cosmological
simulations, we find that the X-ray AGNs reside in halos with the
maximum rotational velocity $\approx 320\,$\kms, or with total masses
$\sim 4.1\times10^{12}\,h^{-1}\,M_{\odot}$, also with no detectable
redshift trend. The lack of a redshift or luminosity dependence of the
AGN clustering is inconsistent with the scenarios in which the AGN
luminosities in the active state are similar fractions of the
Eddington luminosity. However, it can be explained in the scenario in
which the AGN activity is triggered by major mergers of gas-rich
galaxies, and the instantaneous luminosity passes through many levels
after each trigger \citep{2006ApJS..163....1H}.

Our results reveal another interesting aspect of the AGN clustering
which was predicted in \cite{2008ApJS..175..356H}. The redshift
measurements in our sample are sufficiently accurate to detect
peculiar motions of objects in excess of $\sim 100\,$\kms. The
comparison of the two-point correlation functions projected on the
line of sight and on the sky plane reveals no signatures of the
redshift-space distortions, which allows us to put limits on the
fraction of AGNs located in the satellite subhalos \emph{within} the
host dark matter halos. We find that the X-ray AGNs are predominantly
located in the central galaxies of the host dark matter halos and tend
to \emph{avoid} satellite galaxies. Quantitatively, we limit the
fraction of AGNs in non-central galaxies to be $<0.12$ at the 90\% CL.
We also exclude the model in which the probability for a galaxy with
$\vmax>300\,$\kms{} to host an AGN is the same for central galaxies
and satellite galaxies in more massive halos
(\S\,\ref{sec:res:vmin-f}). The central locations of the quasar host
galaxies are expected in the trigger model because mergers of
equally-sized galaxies preferentially occur at the centers of dark
matter halos \citep{2008ApJS..175..356H}.

Finally, we compared the number densities of the \ChBootes{} AGNs to
that of the dark matter halos with the mass corresponding to the AGN
clustering amplitude. We find that the fraction of halos with active
X-ray AGNs decreases with increasing $z$ --- and, correspondingly,
with $L_{x}$ --- probably reflecting a lower probability for an object
to have a higher instantaneous luminosity. At the lowest redshifts in
our sample, \emph{Chandra} probes such low luminosity that X-ray AGNs
become quite common. At $z=0.37$, the \emph{Chandra}-detected sources
are located in more than 10\% of the dark matter halos with
$\vmax>280\,$\kms{} or $M>3.7\times10^{12}\,h^{-1}\,M_{\odot}$.

\acknowledgements

This work was supported by the Smithsonian Institution and \emph{NASA}
through contracts NAS8-39073 and NAS8-03060 (CXC). AK is supported by the
NSF grant AST-0708154, by NASA grant NAG5-13274, and by Kavli
Institute for Cosmological Physics at the University of Chicago
through grant NSF PHY-0551142 and an endowment from the Kavli
Foundation. We thank O.~Gnedin and A.~Franceschini for useful
discussions and comments.

\emph{Facilities:} \emph{Chandra,} MMT.

\***{fillup}

\bibliography{chbootes}

\begin{appendix}

  \section{Cosmology-dependent scalings of the model correlation function}
  \label{sec:cosm-depend-scal}

  Using the simulation outputs directly has many advantages over the
  analytic HOD model in testing different models of populating the
  dark matter halos with astronomical sources. However, there is an
  important disadvantage. The simulations are performed for a certain
  combination of the cosmological parameters, which may or may not
  match the currently favored cosmological model. For example, the
  numerical simulations we use in this work were performed assuming a
  high value of the power spectrum normalization, $\sigma_{8}=0.9$ at
  $z=0$. This results in a slightly incorrect prediction of the
  correlation amplitude for halos of a given mass, and thus slightly
  biases the derived parameters of the AGN population model.

  Fortunately, as we discuss below, the simulation-based models can be
  easily adjusted to the ``correct'' cosmology by simply rescaling the
  halo masses by a small factor. A more elaborate procedure is
  described by \cite{2010MNRAS.405..143A} who show that even the raw
  simulation outputs can be rescaled to the correct cosmology.

  \subsection{Basics}

  The correlation function of objects with mass $M$ at a given
  redshift, $z$, can be written as
  \begin{equation}
    \xi(r,M) = b^{2}(M)\,\xi_{\rm DM}(r),
  \end{equation}
  where $b(M)$ is the bias factor for halos with mass $M$, and
  $\xi_{\rm DM}(r)$ is the correlation function of all dark matter
  particles. Since we consider the correlation function measurements
  at the separations bracketing $8\,h^{-1}\,$Mpc, we can assume that
  with sufficient accuracy, $\xi_{\rm DM}(r)\propto \sigma_{8}^{2}(z) =
  \sigma_{8,0}^{2}D(z)^{2}$, where $D(z)$ , is the linear perturbations growth
  factor for the given cosmology.

  The bias factor, $b(M)$, is, in turn, a function of the linear
  perturbation amplitude at the mass scale $M$, $b(M)=b(\sigma(M))$
  \citep{1984ApJ...284L...9K,1996MNRAS.282..347M,1999MNRAS.308..119S}. Therefore,
  we conclude that the dependence of the correlation function model on
  cosmology is through $\sigma_{8}$, $D(z)$, and $\sigma(M)$.

  \subsection{Scaling for $\sigma(M)$}

  The scales corresponding to the galaxy-sized objects are
  sufficiently small that the slope of the matter power spectrum
  and hence the mass dependence in $\sigma(M)$ is insensitive to the
  underlying cosmology \citep{1986ApJ...304...15B}. In a fairly broad
  range of parameters around the ``concordance'' cosmological model,
  we find that $\sigma(M)$ computed with the full transfer function
  \citep[e.g.,][]{1998ApJ...496..605E} can be approximated as $\sigma(M)=A\times
  M^{-0.16}$ in the mass range $10^{12}-10^{13}\,M_{\odot}$. The
  amplitude, $A$, of this power law approximation scales with
  cosmology as
  \begin{equation}\label{eq:sigma(M):scaling}
    A \propto \sigma_{8} \times D(z) \times s(\Omega_{M}, h, ...).
  \end{equation}
  In this equation, the factor $s(\Omega_{M},h, ...)$ represents the
  correction for the different power spectrum shapes in different
  cosmologies \citep[it is mostly a function of the product
  $\Omega_{M}h$, see][]{1984ApJ...285L..45B}. We neglected such a
  factor in $\xi_{DM}(r)$ because we consider the cosmological models
  in which the perturbations are normalized by $\sigma_{8}$, the
  fluctuation amplitude at approximately the midpoint of the observed
  range. However, the scale corresponding to galaxy-sized objects for
  which $\sigma(M)$ has to be computed, is sufficiently far from
  $8\,h^{-1}\,$Mpc so that the correction can be important. For
  example, we find that $s_{1}/s_{2}=1.03$ for the flat $\Lambda$CDM
  cosmologies with $(\Omega_{M},h)=(0.3,0.7)$ and $(0.268,0.71)$ (more
  power in the high-$\Omega_{M}$ case).

\begin{figure*}
\vspace*{-7mm}
\centerline{
\includegraphics[width=0.49\columnwidth]{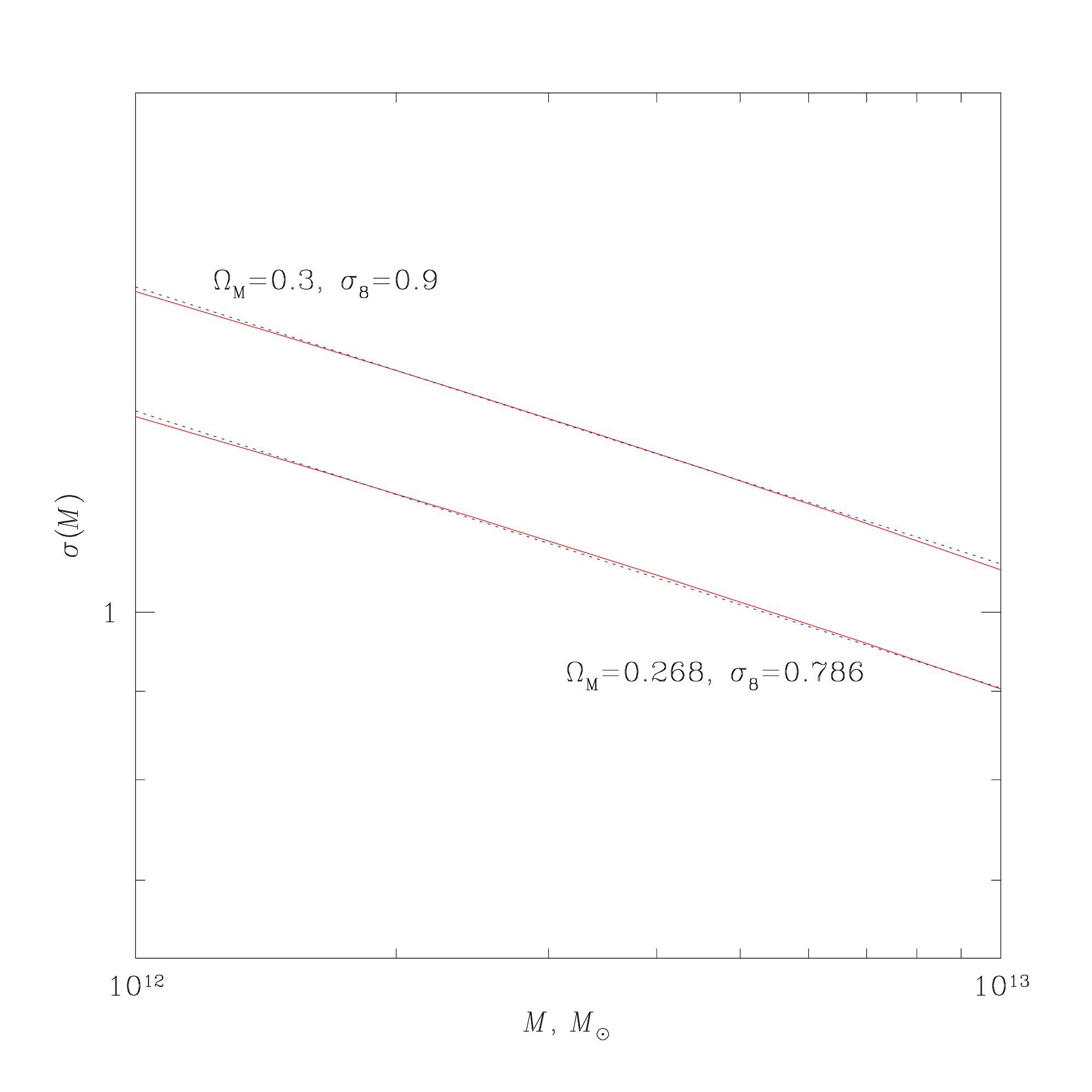}
\includegraphics[width=0.49\columnwidth]{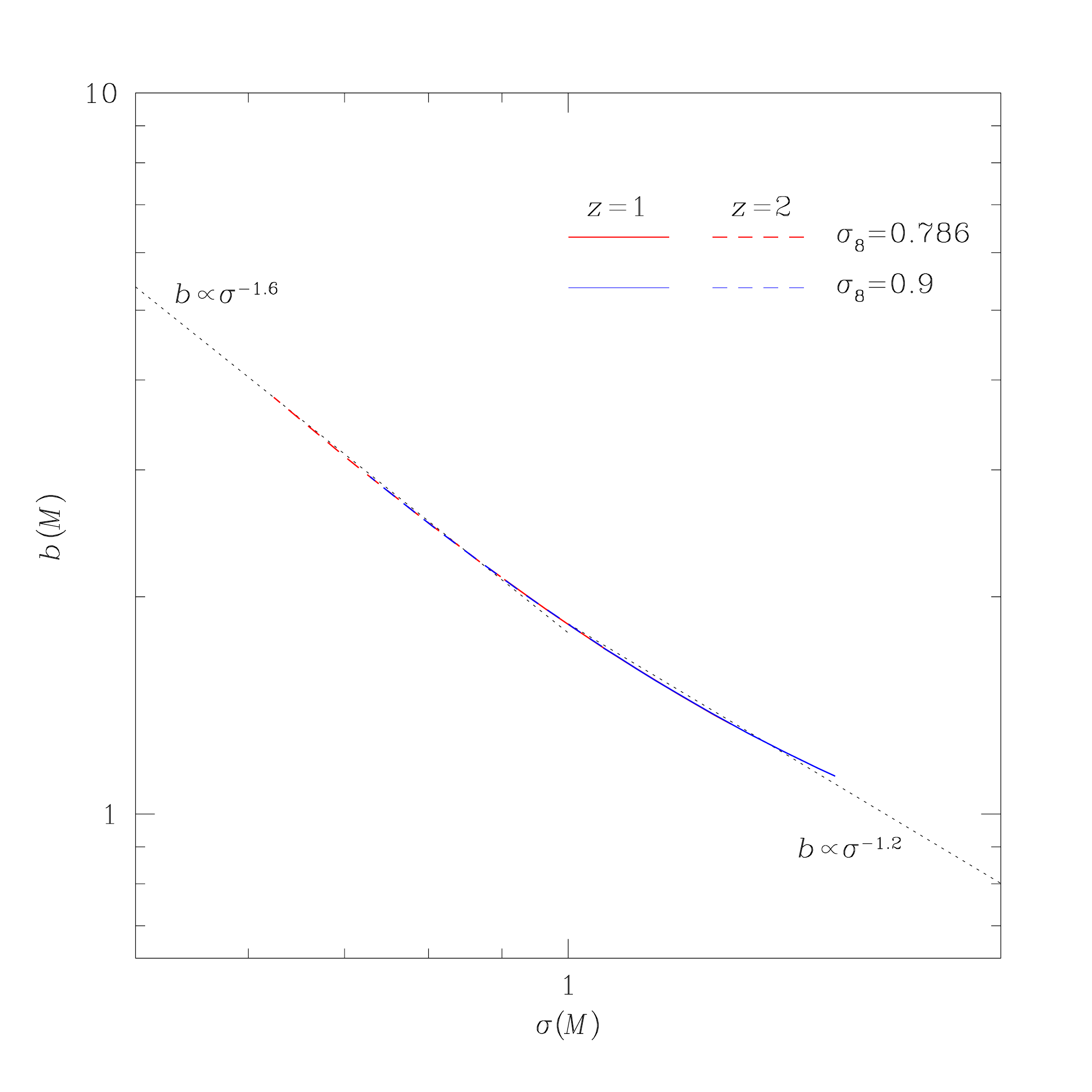}
}
\vspace*{-3mm}
\caption{Cosmological functions $\sigma(M)$ and $b(M)$ computed for
  two different flat $\Lambda$CDM cosmologies,
  $(\Omega_{M},\sigma_{8})=(0.3,0.9)$ and $(0.268,0.786)$. The left
  panel shows $\sigma(M)$ computed at $z=1$ (solid lines) and the
  power law approximations, $\sigma\propto M^{-0.16}$ (dotted
  lines). In the right panel, the linear bias, $b(M)$, is plotted as a
  function of $\sigma(M)$ for the same two cosmologies and for $z=1$
  and 2 (the mass range is $M=10^{12}-10^{13}$ as in the left panel).}
\label{fig:bias:scaling}
\end{figure*}

  \subsection{Scaling for $b(\sigma(M))$}

  The key notion for our computation is that the halos with
  $M=10^{12}-10^{13}$ are ``rare'' objects at $z\gtrsim 1$
  ($\sigma(M)\sim 1$ or less) with high bias factors strongly
  dependent on $\sigma(M)$. The bias as a function of $\sigma(M)$ can
  be computed using the \cite{1999MNRAS.308..119S} approximation:
  \begin{equation}\label{eq:bias:ST}
    b = 1 + \frac{a\,\delta_{c}^{2}/\sigma^{2}-1}{\delta_{c}} 
    + \frac{2p}{\delta_{c}\left[1+\left(a\,\delta_{c}^{2}/\sigma^{2}\right)^{p}\right]}
  \end{equation}
  with parameters $a=0.75$ and $p=0.3$ \citep[adopted from][see
  their page 704]{2003ApJ...584..702H}, and $\delta_{c}=1.69$ is the
  threshold for spherical collapse in a matter-dominated universe. The
  bias computed from this equation for the two cosmologies and two
  different redshifts is shown in the right panel of
  Fig.~\ref{fig:bias:scaling}. For the mass range
  $10^{12}-10^{13}\,M_{\odot}$, the bias can be approximated as
  $b\propto \sigma^{-1.6}$ at $z=2$ and $b\propto \sigma^{-1.2}$ at
  $z=1$.

  \subsection{Scaling of Mass Derived from the Two-Point Correlation Functions}

  Since $\sigma(M)$ is a weak function of mass, the linear bias is
  also a weak function of $M$, $b(M)\propto M^{0.2-0.25}$ in our mass
  range. However, $\sigma(M)$ scales linearly with $\sigma_{8}$, and
  the bias also shows a strong dependence on this parameter, $b\propto
  \sigma_{8}^{-1.6 \ldots -1.2}$. For the correlation function of
  halos, we have $\xi=\xi_{\rm DM}\, b^{2}\propto \sigma_{8}^{-1.2}
  \ldots \sigma_{8}^{-0.4}$. We, therefore, arrive at a somewhat
  counterintuitive conclusion --- the amplitude of the correlation
  function of galaxy-sized objects at high $z$ is \emph{lower} for
  models with high values of $\sigma_{8}$ because of a strong
  dependence of the linear bias on the underlying amplitude of the density
  perturbations, $\sigma(M)$.

  When we determine the mass scale of objects from their correlation
  function, we effectively solve the equation
  \begin{equation}\label{eq:match:obs:xi}
    \xi_{DM}\,b^{2}(M)\propto \sigma_{8}^{2}\, D(z)^{2}\,b^{2}(M)=C,
  \end{equation}
  where $C$ is a constant provided by the data. Inserting the scalings
  derived above, we can rewrite this as
  \begin{equation}
    \sigma_{8}^{2}\, D(z)^{2}\, \left[\sigma_{8}\,D(z)\,s\,M^{-0.16}\right]^{-2\alpha}=C
  \end{equation}
  where we assume that $b(M)$ is approximated as a power law of
  $\sigma(M)$, $b\propto \sigma^{-\alpha}$ (e.g., $\alpha=-1.6$ for
  our mass range and $z=2$). From here we have,
  \begin{equation}\label{eq:mass:scaling}
    M^{0.16} \propto \left[\sigma_{8}\, D(z)\right]^{\alpha-1}\, s^{\alpha}
  \end{equation}
  If we scale the object masses by this equation, the halo clustering
  properties in the two different cosmologies will be very similar.

  As an example, consider the scaling between the cosmology used in
  the simulations, $\Omega_{M}=0.3$, $h=0.7$, $\sigma_{8}=0.9$, and
  the galaxy cluster-normalized best-fit model from
  \cite{2009ApJ...692.1060V}, $\Omega_{M}=0.268$, $h=0.715$,
  $\sigma_{8}=0.786$. At $z=1.03$, the median redshift of our
  \ChBootes{} X-ray sample of AGNs, $D(z)=0.616$ and $0.604$ for
  $\Omega_{M}=0.268$ and 0.3, respectively. The shape factors, $s$,
  are 1 and 1.03 for $\Omega_{M}=0.268$ and 0.3, respectively.
  Inserting into eq.~\ref{eq:mass:scaling}, we find that the masses
  which would be derived in the $\Omega_{M}=0.268$, $\sigma_{8}=0.786$
  cosmology are a factor of 1.45 lower than those derived using the
  correlation function models obtained directly from the simulation
  outputs. 

  At $z=2$, $D(z)=0.284$ and $0.277$ for $\Omega_{M}=0.268$ and 0.3,
  respectively. Using eq.~[\ref{eq:mass:scaling}], we find that the
  object masses should be scaled down by factors $1.7\ldots2$ for
  $\alpha=1.4\ldots1.6$ (the range of slopes for the $z=2$ objects,
  see Fig.~\ref{fig:bias:scaling}).

  \subsection{Scaling for the Number Density of Objects}

  A related question is how to scale the \emph{number density} of
  objects with the given clustering properties. In many situations, it
  should be possible to use a mass function model \citep[e.g., the one
  from][]{1999MNRAS.308..119S} for such a scaling. However, for
  galaxy-sized objects it is possible to derive a simple scaling for
  the number density.

  \cite{2001MNRAS.321..372J} show that the mass function of halos can
  be written in a standard form,
  \begin{equation}\label{eq:mass:function}
    \frac{dn}{dM} = B\times\frac{1}{M^{2}}\frac{d\ln \sigma^{-1}}{d\ln M}\varphi(\sigma),
  \end{equation}
  where $B$ is a constant, $\sigma(M)$ is the linear perturbations
  amplitude on scale $M$, and $\varphi(\sigma)$ is a ``universal''
  function. For the \cite{1999MNRAS.308..119S} model,
  \begin{equation} \varphi(\sigma) = 0.32 \sqrt{\frac{2a}{\pi}}
    \left[1+\left(\frac{\sigma^{2}}{a\,\delta_{c}^{2}}\right)^{p}\right]
    \,\frac{\delta_{c}}{\sigma}\,
    \exp\left(-\frac{a\,\delta_{c}^{2}}{2\,\sigma^{2}}\right)
  \end{equation} where the parameters $a$ and $p$ are the same as in
  the expression for the linear bias (eq.~[\ref{eq:bias:ST}]). For
  $\sigma\lesssim1$, as is the case for our galaxy-sized objects, the
  Sheth \& Tormen $f$ is a slowly varying function of $\sigma$ which can
  be approximated as
  \begin{equation}
    \varphi(\sigma)\simeq 0.65 + \ln\sigma
  \end{equation}
  Inserting this into eq.~[\ref{eq:mass:function}] and
  taking into account that $\sigma(M)\propto M^{-0.16}$ (see above), we
  have for the number density of objects above a mass threshold $\Mlim$,
  \begin{equation}\label{eq:N:scaling}
    N = \int_{\Mlim}^{\infty}
    B\times\frac{0.16}{M^{2}}\left[0.65+\ln\sigma_{\rm
        lim}-0.16\ln(M/\Mlim)\right]\,dM \propto
    \Mlim^{-1}\,(0.49+\ln\sigma_{\rm lim})
  \end{equation}
  where $\sigma_{\rm lim} = \sigma(\Mlim)$. Now we can
  take into account that the limiting mass is found from the
  condition~\ref{eq:match:obs:xi} (so that the clustering properties
  match those observed). If bias is approximated as a power law of
  $\sigma$, $b(M)\propto \sigma(M)^{-\alpha}$ (Fig.~\ref{fig:bias:scaling}),
  \begin{equation}\label{eq:sigmalim:scaling}
    \sigma_{\rm lim}\propto
    \left[\sigma_{8}D(z)\right]^{1/\alpha}.
  \end{equation}
  Equations~\ref{eq:N:scaling},~\ref{eq:sigmalim:scaling},
  and~\ref{eq:mass:scaling} provide the scaling for the number density
  of isolated halos whose clustering properties match the
  observations. Assuming that the relative number density of satellite
  and main halos is not very sensitive to the underlying cosmology,
  the same scaling can be applied to any other type of halos.

  \bigskip

  In summary, we suggest the following procedure. One takes the
  simulation outputs and adjusts the mass threshold, $\Mlim$ so that
  the correlation function amplitude derived for the halos with mass
  above this threshold matches that observed. One then measures the
  number density, $N$, of such halos in the simulation box. To convert
  these quantities to the desired cosmology, one uses
  eq.\,\ref{eq:mass:scaling} to scale $\Mlim$ and equations
  \ref{eq:N:scaling},~\ref{eq:sigmalim:scaling},
  and~\ref{eq:mass:scaling} for the number density. For example, we
  found for the \ChBootes{} AGNs $\Vmin=320\,$\kms, which corresponds
  to $\Mlim=4.1\times10^{12}\,h^{-1}\,M_{\odot}$. This corresponds to
  $\sigma_{\rm lim}=1.25$ at $z=1$. To adjust these results from the
  $(\Omega_{M},\sigma_{8})=(0.3,0.9)$ to $(0.268,0.786)$ cosmology, we
  need to scale $\Mlim$ by a factor of 0.69, and increase the number
  density by a factor of 1.2 (the $\Mlim^{-1}$ factor is partially
  compensated for by the change in $\ln\sigma_{\rm lim}$).

\end{appendix}

\end{document}